\documentclass[%
 aip,
 amsmath,amssymb,
 reprint,%
]{revtex4-2}

\usepackage{graphicx}
\usepackage{dcolumn}
\usepackage{bm}
\usepackage[utf8]{inputenc}
\usepackage[T1]{fontenc}
\usepackage{mathptmx}
\usepackage{etoolbox}
\usepackage{xcolor}

\makeatletter
\def\@email#1#2{%
 \endgroup
 \patchcmd{\titleblock@produce}
  {\frontmatter@RRAPformat}
  {\frontmatter@RRAPformat{\produce@RRAP{*#1\href{mailto:#2}{#2}}}\frontmatter@RRAPformat}
  {}{}
}%
\makeatother

\begin{document}

\preprint{AIP/123-QED}

\title{Transport of a self-propelled tracer through a hairy cylindrical channel: interplay of stickiness and activity}
% Force line breaks with \\
\author{Rajiblochan Sahoo}
 \affiliation{Department of Chemistry, Indian Institute of Technology Bombay, Mumbai, Maharashtra -  400076, India}
\author{Ligesh Theeyancheri}
\affiliation{Department of Chemistry, Indian Institute of Technology Bombay, Mumbai, Maharashtra -  400076, India}
\author{Rajarshi Chakrabarti$^\ast$}
\email{rajarshi@chem.iitb.ac.in}
\affiliation{Department of Chemistry, Indian Institute of Technology Bombay, Mumbai, Maharashtra -  400076, India}

\begin{abstract}
\noindent Active transport of biomolecules assisted by motor proteins is imperative for the proper functioning of cellular activities. Inspired by the diffusion of active agents in crowded cellular channels, we computationally investigate the transport of an active tracer through a polymer grafted cylindrical channel by varying the activity of the tracer and stickiness of the tracer to the polymers. Our results reveal that the passive tracer exhibits profound subdiffusion with increasing stickiness by exploring deep into the grafted polymeric zone, while purely repulsive one prefers to diffuse through the pore-like space created along the cylindrical axis of the channel. In contrast, the active tracer shows faster dynamics and intermediate superdiffusion even though the tracer preferentially stays close to the dense polymeric region. This observation is further supported by the sharp peaks in the density profile of the probability of radial displacement of the tracer. We discover that the activity plays an important role in deciding the pathway that the tracer takes through the narrow channel. Interestingly, increasing the activity washes out the effect of stickiness. Adding to this, van-Hove functions manifest that the active tracer dynamics deviates from Gaussianity, and the degree of deviation grows with the activity. Our work has direct implications on how effective transportation and delivery of cargo can be achieved through a confined medium where activity, interactions, and crowding are interplaying. Looking ahead, these factors will be crucial for understanding the mechanism of artificial self-powered machines navigating through the cellular channels and performing \textit {in vivo} challenging tasks.
\end{abstract}

\maketitle

\section{Introduction}\label{Intro}

\noindent The diffusion of macromolecules in complex crowded environments plays a significant role in the smooth functioning of living cells. For example, a typical physiological media like cytoplasm inside the cell contains a variety of macromolecules such as proteins, enzymes, DNA, etc~\cite{mcguffee2010diffusion,chapman2014onset}. The diffusion of these biomolecules plays important roles in different biochemical processes such as protein-protein association, enzyme reactions~\cite{berry2002monte}, gene transcription~\cite{wang2012disordered} \textit{etc}. The crowding associated with the environment greatly influences the mechanism of \textit{in vivo} and \textit{in vitro} molecular diffusions, for example nanoparticle diffusion in the cytoplasmic fluid of living cells\cite{norregaard2017manipulation}, intracellular transport, and proteins diffusing through the mucus membrane\citep{tabei2013intracellular} or nuclear pore complex (NPC), which serves as a gateway connecting the nucleoplasm and cytoplasm of cells~\cite{chatterjee2011subdiffusion,chakrabarti2013tracer,chakrabarti2014diffusion,patel2007natively,bickel2002nuclear}. \\

\noindent Over the past decades, the diffusion of probe particles in crowded media has been widely studied both experimentally~\cite{herrmann2009near,pederson2000diffusional,wang2012disordered,devetter2014observation,du2019study,kumar2013nanocomposites,kohli2012diffusion,lee2017nanorod,garamella2020anomalous} and theoretically~\cite{samanta2016tracer,kumar2019transport,yuan2019activity,kaiser2020directing,wu2021mechanisms,sorichetti2021dynamics,debets2020characterising,metzler2014anomalous,ghosh2015non,godec2014collective}. However, most of these studies focus on the passive diffusion of particles in crowded heterogeneous environments, but in the context of cellular biology, there are plenty of examples of diffusion of active particles such as molecular motors~\cite{zheng2000prestin,duan2016ubiquitin,sundararajan2008catalytic}, active filaments~\cite{loose2014bacterial}, microtubules~\cite{sumino2012large} \textit{etc}. More recently, researchers have come up with artificial microswimmers like self-propelled Janus particles or chiral particles mimicking the role of biological swimmers\cite{lozano2018run,gao2015artificial,volpe2011microswimmers}. In recent years, dynamics of self-propelled agents in viscoelastic and crowded environments have been studied by experimentalists as well as investigated in computer simulation. For example, experimental investigation of self-propelled Janus particles in viscoelastic fluid demonstrated that the rotational diffusion of the particle gets enhanced with the self-propulsion velocity~\cite{gomez2016dynamics}. Computer simulations have shown that the rotational diffusivity of a self-propelled Janus particle in the crowded environment exhibits a nonmonotonous behavior while translational diffusivity decreases monotonically with the area fraction of the crowders \cite{theeyancheri2020translational,abaurrea2020autonomously}. On the other hand, enhancement in translational motion of the cell and a sharp decline in rotational diffusion are observed in an experimental study of E. \textit{coli} in polymeric solutions\cite{patteson2015running}. The dynamical behavior is significantly affected by the activity, degree of confinement, steric hindrance, interactions with the medium, and architecture of the surrounding environment. A more fundamental question would be to ask how these factors influence biological processes. A growing interest lies in understanding the transport of macromolecules through cellular channels~\cite{chakrabarti2013tracer}, which is different from studying the particle diffusing in a typical crowded environment as here, other than the crowding, confinement effects are also important. In addition, a key feature of these channels is selectivity and specificity. Biological cells use a wide variety of transmembrane channels and pumps acting as selective barriers for the passage of different proteins or biomolecules~\cite{nakielny1999transport,bruckbauer2010selective}. Among the cellular channels, the most celebrated one is the nuclear pore complex, which regulates the nuclear import and export between the nucleus and the cytoplasm~\cite{shahin2016gatekeepers,dingwall1982polypeptide}. NPCs allow the passive diffusion of ions and small molecules through the narrow aqueous channels. This bidirectional transport is an essential process in eukaryotic cells. NPCs control the movement of various proteins or biomolecules in and out of the nucleus based on the size~\cite{whittaker2000viral}. This size-selective transport of large molecules through NPC were reported earlier~\cite{caspi2008synthetic,fragasso2021designer}. The knowledge of the functioning and transport mechanism of the particles through the membrane channels like NPC has implications in cellular processes. For example, during viral infections, the virus crosses the NPC channel and replicates inside the nucleus in eukaryotic cells~\cite{whittaker2000viral,zila2021cone}. So the mechanism underlying the transport of particles through cellular channels is essential for the smooth functioning of living organisms. However, these are mostly the cases of passive transport. It will be interesting to explore the dynamics of active tracers, representative of synthetic nanomotors in crowded narrow channels. \\

\noindent In this work, we investigate the transport of a self-propelled particle through a cylindrical channel grafted with polymers from inside. This serves as the minimalistic computer model for a cellular channel like NPC, which regulates the molecular diffusion in and out of the nucleus. We extensively analyze the dynamics of the self-propelled particle in this crowded cylindrical channel. Using computer simulations, we investigate the effect of excluded volume, short-ranged sticky interactions, and activity on the dynamics of a self-propelled particle. In general, we find that the dynamics of the probe particle is always enhanced with activity. However, even though the dynamics is faster with increasing activity, there is no preferred direction of transport. The active tracer frequently changes its direction while translating due to the presence of crowders (polymer chains), in addition to thermal noise. When the polymers are sticky to the tracer, the tracer is pulled inside, be it passive or active. In particular, when the activity is high, the tracer is to be found everywhere and the strength of stickiness does not matter. This also reveals that for the passive tracer, the radial movement is prominent in the case of attractive interaction while most of the diffusion occurs along the cylindrical axis of the channel for the repulsive case. Thus interaction strength also plays a vital role here. This suggests that a combination of moderate activity and stickiness is a preferred choice for efficient transport and targeted delivery. Higher activity destroys specificity and higher stickiness makes the diffusion process slow and inefficient. On the other hand, low stickiness (or absence of it) does not help the tracer to find its target buried inside the polymeric zone. Understanding the mechanism of active transport has practical relevance, in the context of targeted delivery such as motor-based or pump-based drug delivery using self-propelled nanotransporters~\cite{patra2013intelligent}. The transportation and delivery of cargo at specific locations using self-powered agents can be an efficient tool for biomedical applications. Hence, studies of a self-propelled particle diffusing through a narrow channel will help in designing realistic experiments to explain the mechanism underlying the biological motor-based transport through cellular channels.

\section{Model and simulation details}\label{Model}

\begin{figure}[h]
\centering
 \begin{tabular}{c}
  \includegraphics[width=0.99\linewidth]{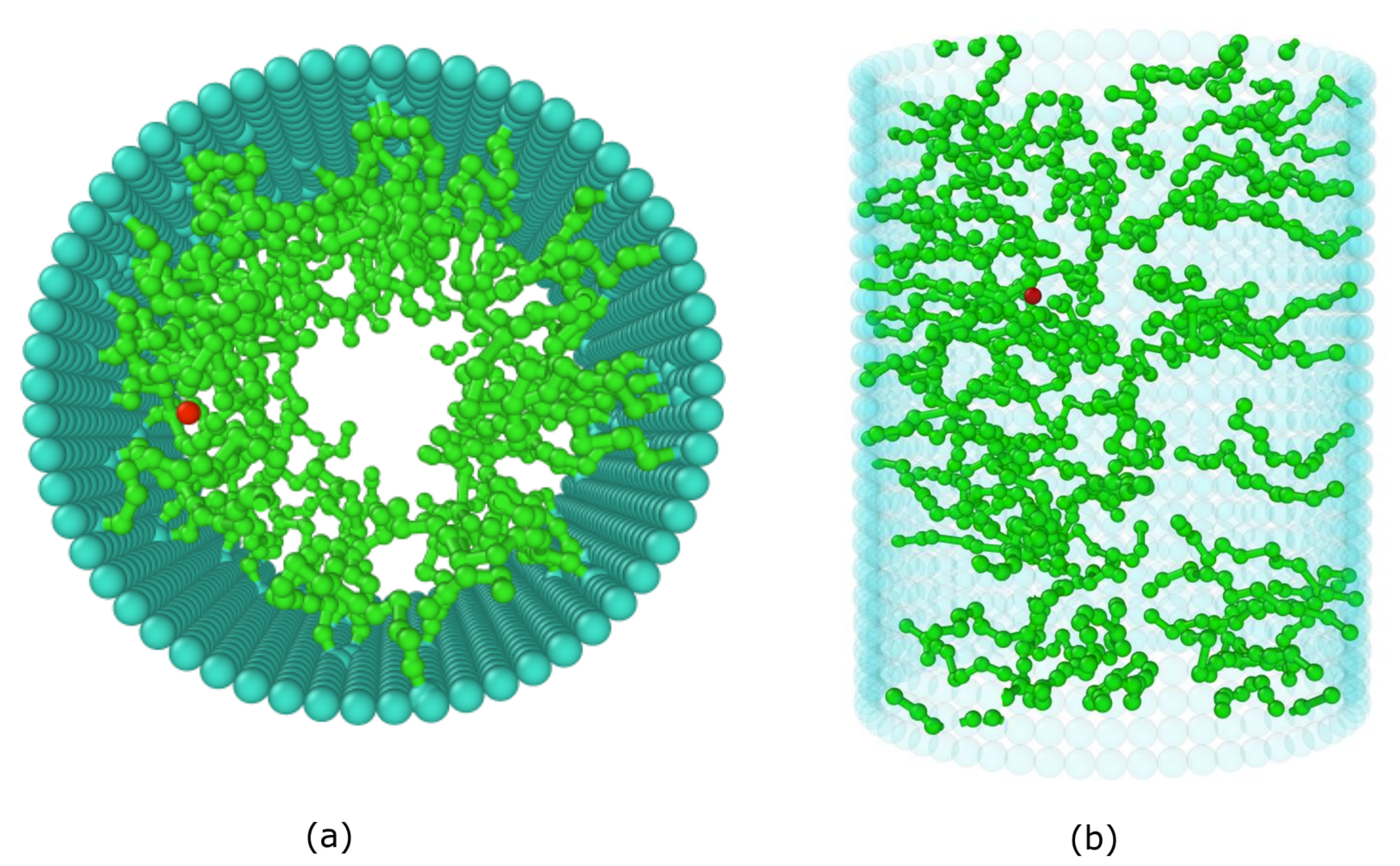}
  \end{tabular}
  \caption{(a) Top view and (b) side view of the snapshot of a representative tracer particle (red) inside the polymer (green) grafted cylindrical channel with rigid walls (cyan). The wall particles are made transparent in the side view to show the grafted polymers clearly. The snapshot is generated using the visualization package OVITO~\cite{stukowski2009visualization}.  Here the size of the tracer is the same as that of a monomer}.
  \label{fig:model}
\end{figure}

\noindent We model the hairy narrow cylindrical channel by grafting one terminal monomer of the linear polymers to one of the inner wall particles of a cylinder with height $24 \sigma$ and radius $9\sigma$ in three dimensions with periodic boundary conditions along the cylinder axis (Fig.~\ref{fig:model}). The cylinder axis is chosen to be the $z$ axis. The cylinder is static throughout the simulations. A total of 75 polymer chains are grafted and each of the polymer chains consists of 12 monomers, connected to the neighboring monomers by finite extensible nonlinear elastic (FENE) potential,

\begin{equation}
V_{\text{FENE}}\left(r \right)=\begin{cases} -\frac{k r_{\text{max}}^2}{2} \ln\left[1-\left( {\frac{r_{ij}}{r_{\text{max}}}}\right) ^2 \right],\hspace{3mm} \mbox{if } r_{ij} \leq r_{\text{max}}\\
\infty, \hspace{35mm} \mbox{otherwise}.
\end{cases}
\label{eq:FENE}
\end{equation}
where $r_{ij}$ is the distance between two neighboring monomers in the polymer, $r_{\text{max}}$ is the maximum displacement of the bond, and $k$ is the force constant. In our simulation the parameters are $k = 7$, $r_{\text{max}} = 5$. The repulsive interaction between the polymer beads and with the wall particles is modeled via the purely repulsive Weeks-Chandler-Andersen (WCA) potential~\cite{weeks1971role},
\begin{equation}
V_{\textrm{WCA}}(r_{ij})=\begin{cases}4\epsilon_{ij}\left[\left(\frac{\sigma_{ij}}{r_{ij}}\right)^{12}-\left(\frac{\sigma_{ij}}{r_{ij}}\right)^{6}\right]+\epsilon_{ij}, \hspace{5mm} \mbox{if }r_{ij}<2^{1/6}\sigma_{ij} \\
0, \hspace{34mm} \mbox{otherwise},
\end{cases}
\label{eq:WCA}
\end{equation}
where $\sigma_{ij}=\frac{\sigma_{i}+\sigma_{j}}{2}$ is the effective interaction diameter, $r_{ij}$  is the separation between a pair of interacting monomers and $\epsilon_{ij}$ is the strength of the interaction. \\

\noindent We introduce a spherical tracer particle (red color in Fig.~\ref{fig:model}) of diameter $1 \sigma$ inside this cylindrical channel grafted with polymers. All the particles in the system have identical masses. The tracer particle interacts repulsively with the wall particles by WCA potential and with the grafted polymers either $via$ a repulsive WCA potential or $via$ a sticky interaction, modeled by the standard Lennard-Jones potential,
\begin{equation}
V_{\textrm{LJ}}(r_{ij})=\begin{cases}4\epsilon_{ij}\left[\left(\frac{\sigma_{ij}}{r_{ij}}\right)^{12}-\left(\frac{\sigma_{ij}}{r_{ij}}\right)^{6}\right], \hspace{5mm} \mbox{if } r_{ij} \leq r_{\textrm{cut}}\\
0 \hspace{31mm} , \mbox{otherwise}\\
\end{cases}
\label{eq:LJ}
\end{equation}
where $r_{ij}$ is the separation between the tracer particle and monomers of the grafted polymers, $\epsilon_{ij}$ is the strength of the interaction (stickiness) with an interaction diameter $\sigma_{ij}$, and the Lennard-Jones cutoff length $r_{\textrm{cut}}$= $2.5$ $\sigma$. We have varied the interaction strength $\epsilon$ and size of the tracer particle $\sigma_p$ in our simulations. The Lennard-Jones parameters ($\sigma$ and $\epsilon$) and $m$ are the fundamental units of length, energy, and mass, respectively. Therefore, the unit of time is $ \tau = \sqrt{m\sigma^2/\epsilon}$.\\

\noindent The following Langevin equation is implemented to simulate the dynamics of a particle with mass $m$ and position $r_{i}(t)$ at time $t$, interacting with all the other particles in the system. \\
\begin{equation}
m_{i}\frac{d^2 \textbf{r}_{i}(t)}{dt^2}=-\xi \frac{d \textbf{r}_{i}}{dt}- \sum_{j} \bigtriangledown V(\textbf{r}_i-\textbf{r}_j)+ {\bf f}_{i}(t)+{\bf f}_{\text{act}} \bm{n}\\
\label{eq:langevineq}
\end{equation}

\begin{equation}
\frac{d\bm{n}}{dt} = \bm{\eta}(t) \times \bm{n}
\label{eq:orientation}
\end{equation}
here, $r_{j}$ represents the position of all the particles except the $i^{th}$ particle in the system, $V(r) = V_{\text{LJ}}+V_{\text{WCA}}+V_{\text{FENE}}$ is the resultant pair potential between the $i^{th}$ and $j^{th}$ particles. $V_{\text{LJ}} = 0$ for purely repulsive interactions, and we set $V_{\text{WCA}} = 0$ for attractive interactions. Here, we consider very high  friction coefficient  ($\xi = 2.148 \times 10^4$), therefore the dynamics is overdamped. Thermal fluctuations are captured by the Gaussian random force $f_i(t)$, following the fluctuation-dissipation theorem.
\begin{equation}
\left<f_\alpha(t)\right>=0, \hspace{5mm}
\left<f_{\alpha}(t^{\prime})f_{\beta}(t^{\prime\prime})\right> = 6 \xi k_B T \delta_{\alpha\beta}\delta(t^{\prime}-t^{\prime\prime})
\label{eq:random-forcerouse}
\end{equation}
where $k_B$ is the Boltzmann constant and T is the temperature. $\text{f}_{\text{act}}$ represents the magnitude of the active force with orientation specified by the unit vector $\bm{n}$. The orientation changes according to eqn~\ref{eq:orientation}, where $\bm{\eta}(t)$ is the Gaussian distributed stochastic vector with $\left<\bm{\eta}(t)\right>=0$ and time correlations given by $\left<\bm{\eta}_\alpha(t^{\prime})\bm{\eta}_\beta(t^{\prime\prime})\right> = 2 D_r \delta_{\alpha\beta}\delta(t^{\prime}-t^{\prime\prime})$, where $D_r$ is the rotational diffusion coefficient. The unit vector $\bm{n}$ can be expressed in the form of spherical coordinates~\cite{du2019study}. Here we express the self-propulsion in terms of a dimensionless quantity P\'{e}clet number, $\text{Pe}$  defined as $\text{Pe} = \frac{\text{f}_{\text{act}} \sigma}{k_B T}$. Therefore, $\text{Pe} = 0$ corresponds to a passive tracer. \\

\noindent All the simulations are performed using the Langevin thermostat and the equation of motion (eqn ~\ref{eq:langevineq}) is integrated using the velocity Verlet algorithm in each time step. We initialize the system by placing a tracer inside the cylinder grafted with polymer and relaxed the initial configuration for $10^7$ steps. This also ensures the equilibration of the tracer which is placed at the axis of the cylinder in the initial configuration. All the production simulations are carried out for $5 \times 10^8$ steps where the integration time step is considered to be $5 \times 10^{-4}$ and the position of the tracer particle is recorded every $100^{th}$ steps. We have performed $10$ independent simulations for each case. The simulations are carried out using LAMMPS~\cite{plimpton1995fast}, a freely available open-source molecular dynamics package.

\section{Results and discussion}

\begin{figure*}[ht]
\centering
\begin{tabular}{cc}
 \includegraphics[width=0.47\linewidth]{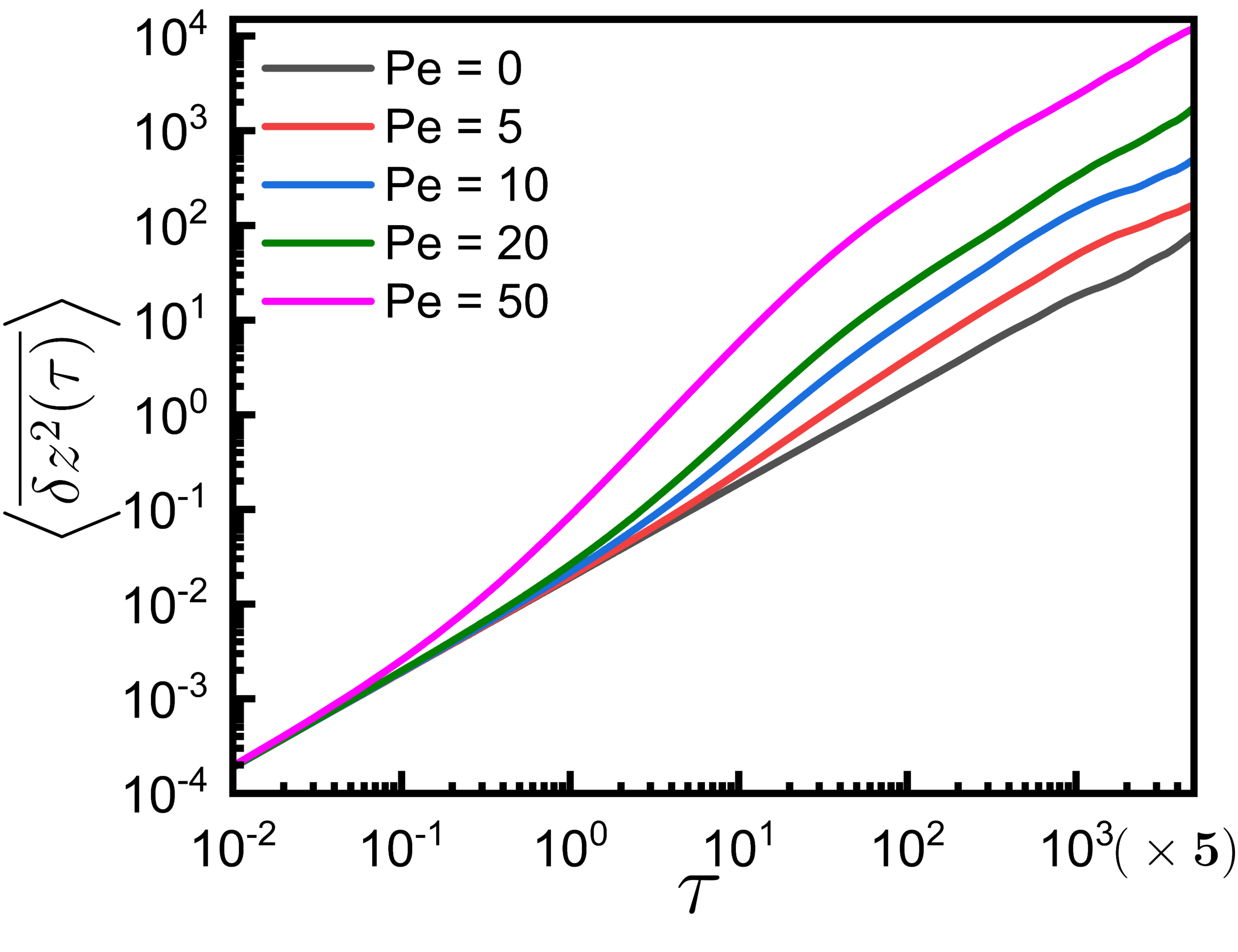} & \includegraphics[width=0.5\linewidth]{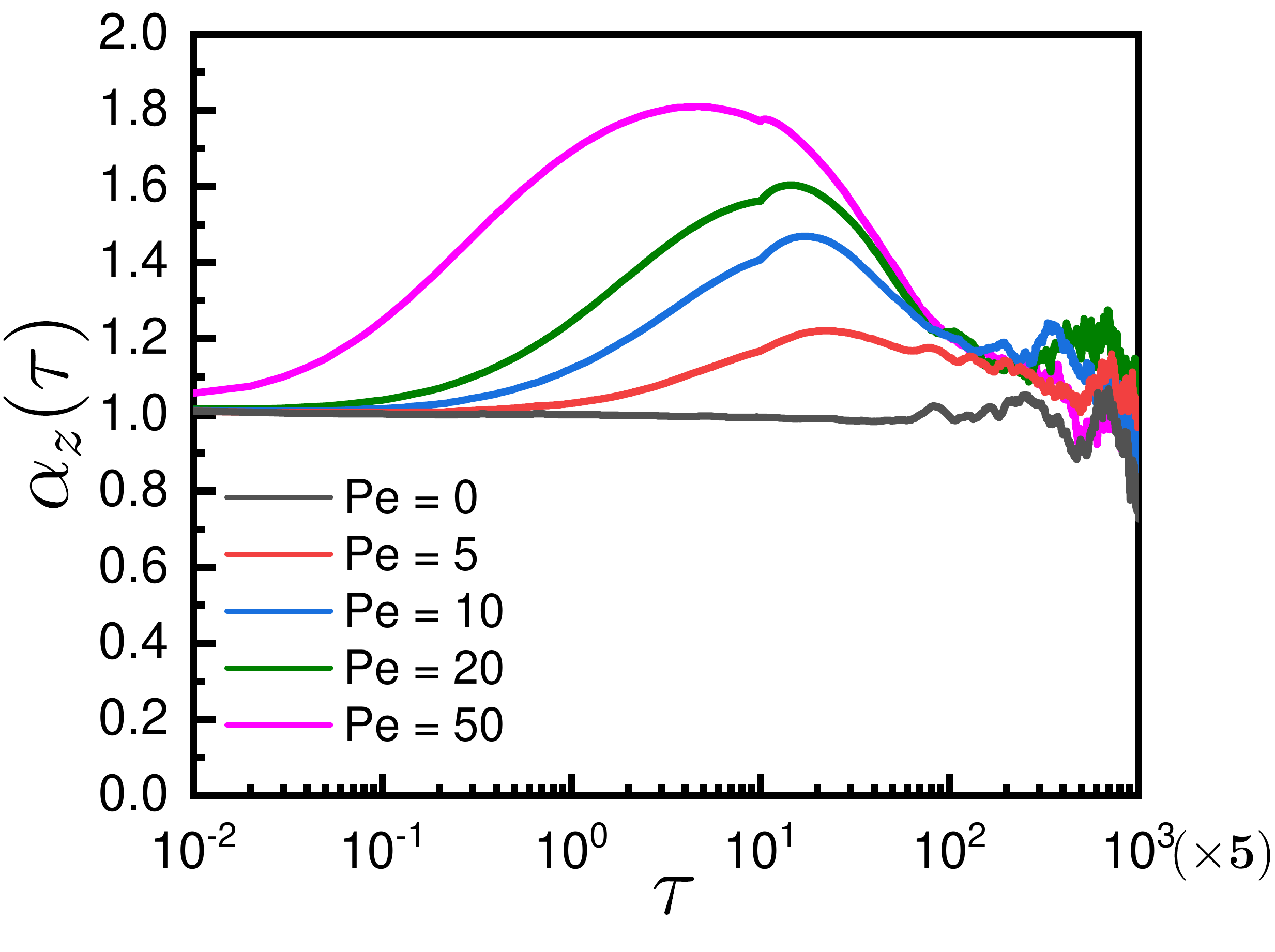} \\
 (a) & (b) \\
 \end{tabular}
 \caption{(a) Log–log plot of $<\delta z^2(\tau)>$ vs $\tau$ and (b) log-linear plot of $\alpha_z(\tau)$ of the tracer particle at different $\text{Pe}$ in cylindrical channel without polymers.}
 \label{fig:free}
\end{figure*}

\noindent As a first step, we simulate the passive $(\text{Pe} = 0)$ and self-propelled $(\text{Pe} \neq 0)$ tracer particle in a free cylindrical channel without incorporating any crowders and study the dynamics for validating the set of parameters used for our work. First we compute the time-averaged MSD, $\overline{\delta{r_i^2}(\tau)} = \frac{1}{T_{\text{max}}-\tau} \int_{0}^{T_{\text{max}}-\tau} {\left[ \textbf{r}_i(t+\tau) - \textbf{r}_i(t)\right]}^2 dt$, from the time evolution of $\textbf{r}_i(t)$, where, $T_{\text{max}}$ is the total run time and $\tau$ is the lag time. Further, to obtain the time-and-ensemble averaged MSD, we calculate the average, $\left\langle\overline{\delta r_{i}^{2}(\tau)}\right\rangle = \frac{1}{N} \sum_{i=1}^N\overline{\delta r_{i}^{2}(\tau)}$, where $N$ is the number of independent trajectories. Here, we compute time-and-ensemble averaged translational mean square displacement (MSD) of the self-propelled tracer particle along the cylinder ($\left\langle\overline{\delta z^{2}(\tau)}\right\rangle$) as well as along the radial ($\left\langle\overline{\delta {(x^2+y^2)}(\tau)}\right\rangle$) directions as a function of lag time $\tau$ for different values of $\text{Pe}$. In the absence of any crowders, the self-propelled tracer particle freely diffuses and the MSDs grow faster in comparison to the passive tracer and exhibit a three-step growth; short time thermal diffusion ($\alpha_z(\tau) = 1$), intermediate superdiffusion ($\alpha_z(\tau) > 1$), and a long time enhanced diffusion ($\alpha_z(\tau) = 1$) (Fig.~\ref{fig:free}), where the time exponent $\alpha_z(\tau) =\frac{d {\text{log}}{(\left\langle\overline{\delta z^{2}(\tau)}\right\rangle)}}{d\text{log}(\tau)}$. On increasing $\text{Pe}$, the growth of $\left\langle\overline{\delta z^{2}(\tau)}\right\rangle$ and $\left\langle\overline{\delta {(x^2+y^2)}(\tau)}\right\rangle$ becomes faster. However, for higher $\text{Pe}$, $\left\langle\overline{\delta {(x^2+y^2)}(\tau)}\right\rangle$ shows a saturation at long time limit due to the confinement created by the wall particles of the cylindrical channel (Fig.~S1). This observation is absent for smaller values of $\text{Pe}$, as the tracer particle is not able to reach the wall of the cylinder to feel the confinement. \\

\begin{figure*}[t]
\centering
\begin{tabular}{cc}
 \includegraphics[width=0.46\linewidth]{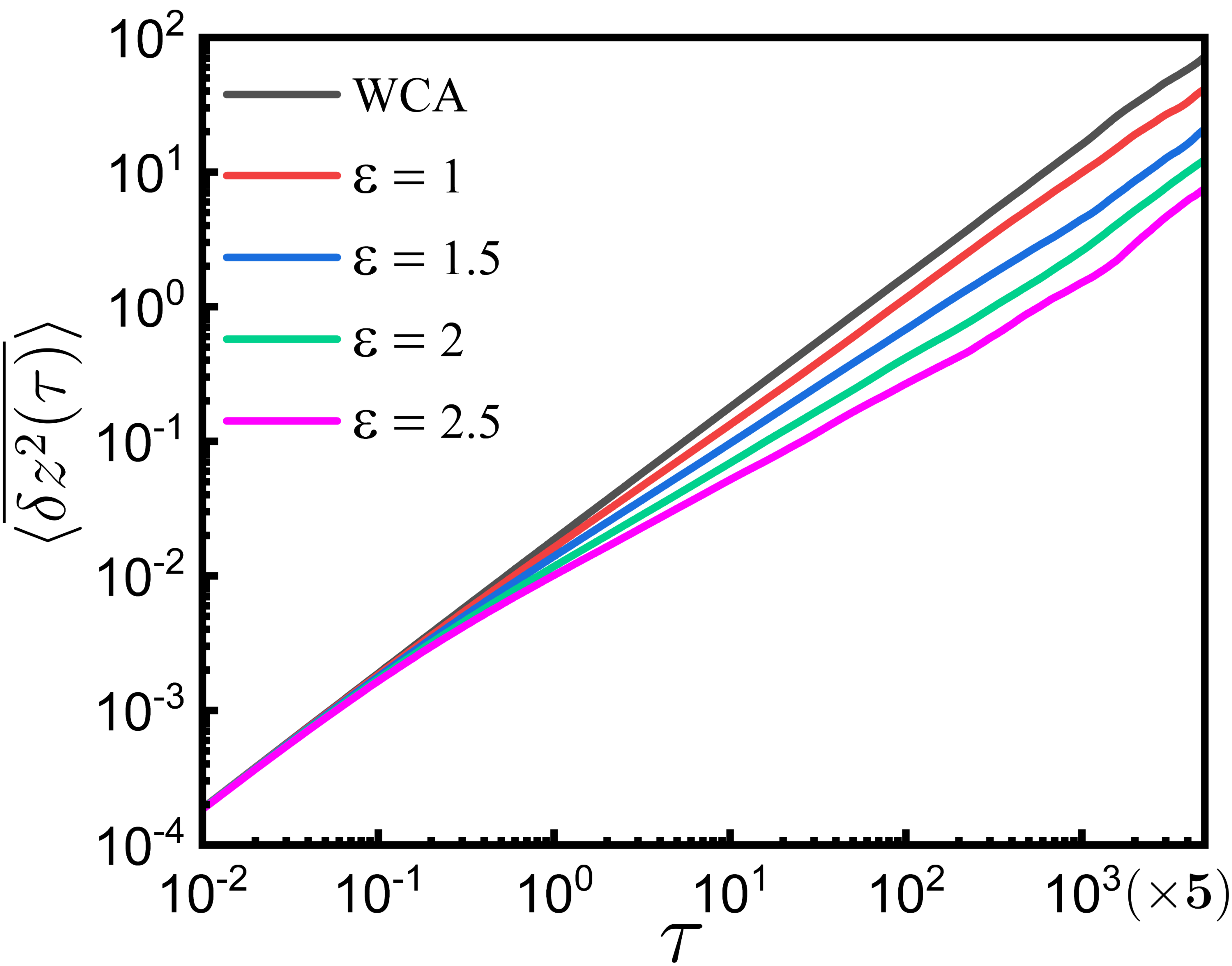} & \includegraphics[width=0.5\linewidth]{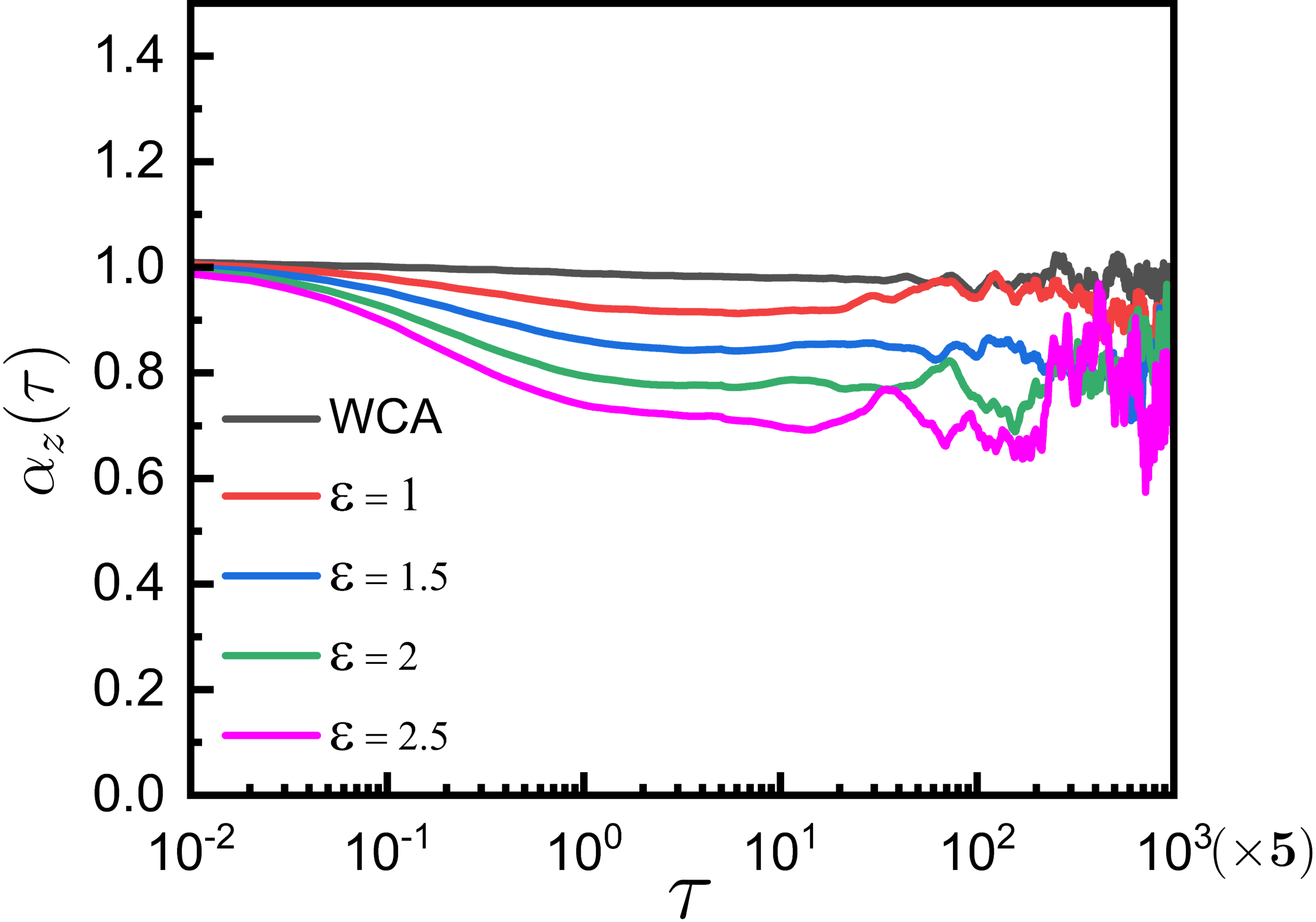} \\
 (a) & (b) \\
 \end{tabular}
 \caption{(a) Log–log plot of $<\delta z^2(\tau)>$ vs $\tau$ and (b) log-linear plot of $\alpha_z(\tau)$ of the passive tracer particle with different stickiness parameter ($\epsilon$) in the polymer grafted cylindrical channel.}
 \label{fig:figure1}
\end{figure*} 

\begin{figure*}[t]
\centering
\begin{tabular}{cc}
 \includegraphics[width=0.46\linewidth]{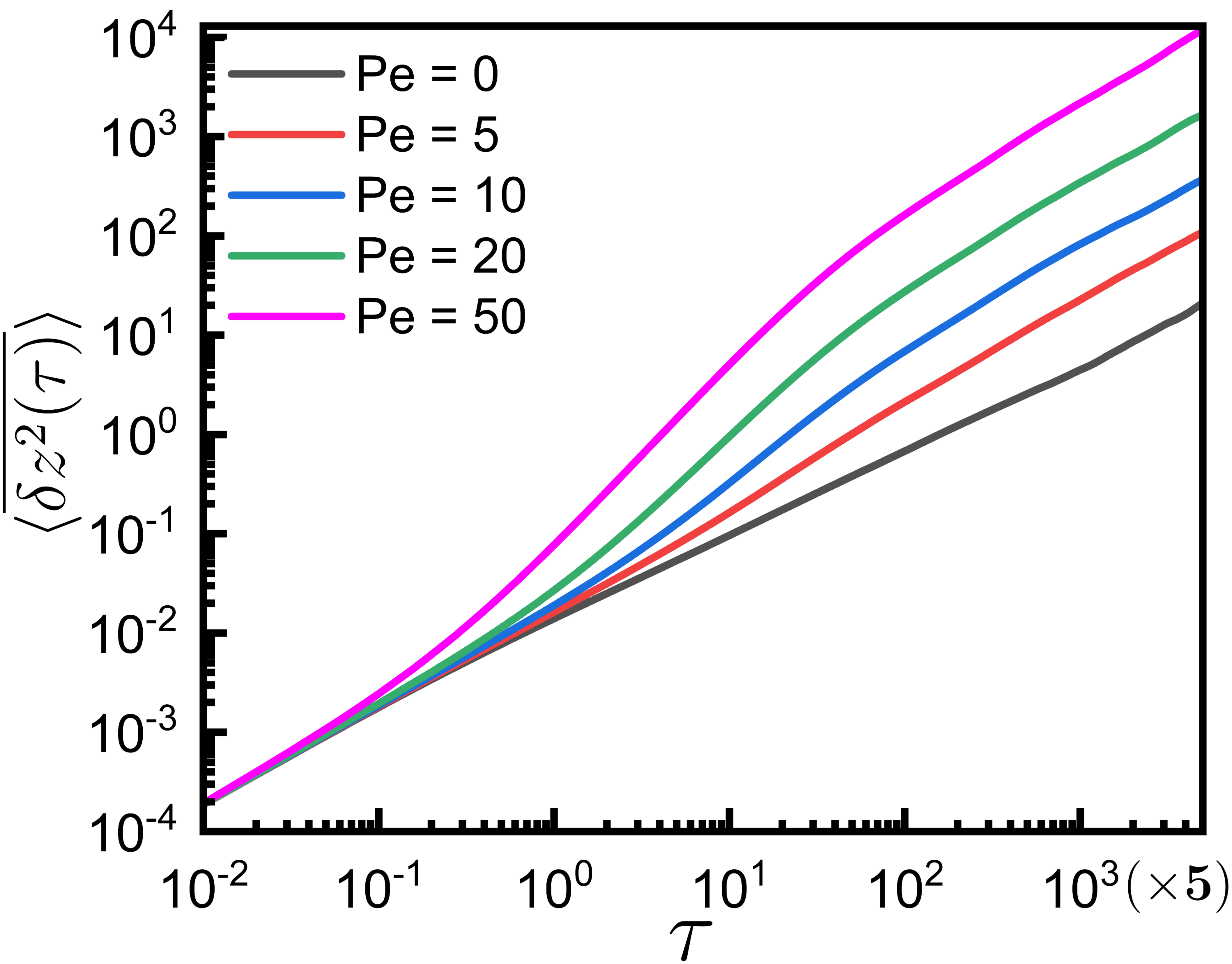} & \includegraphics[width=0.5\linewidth]{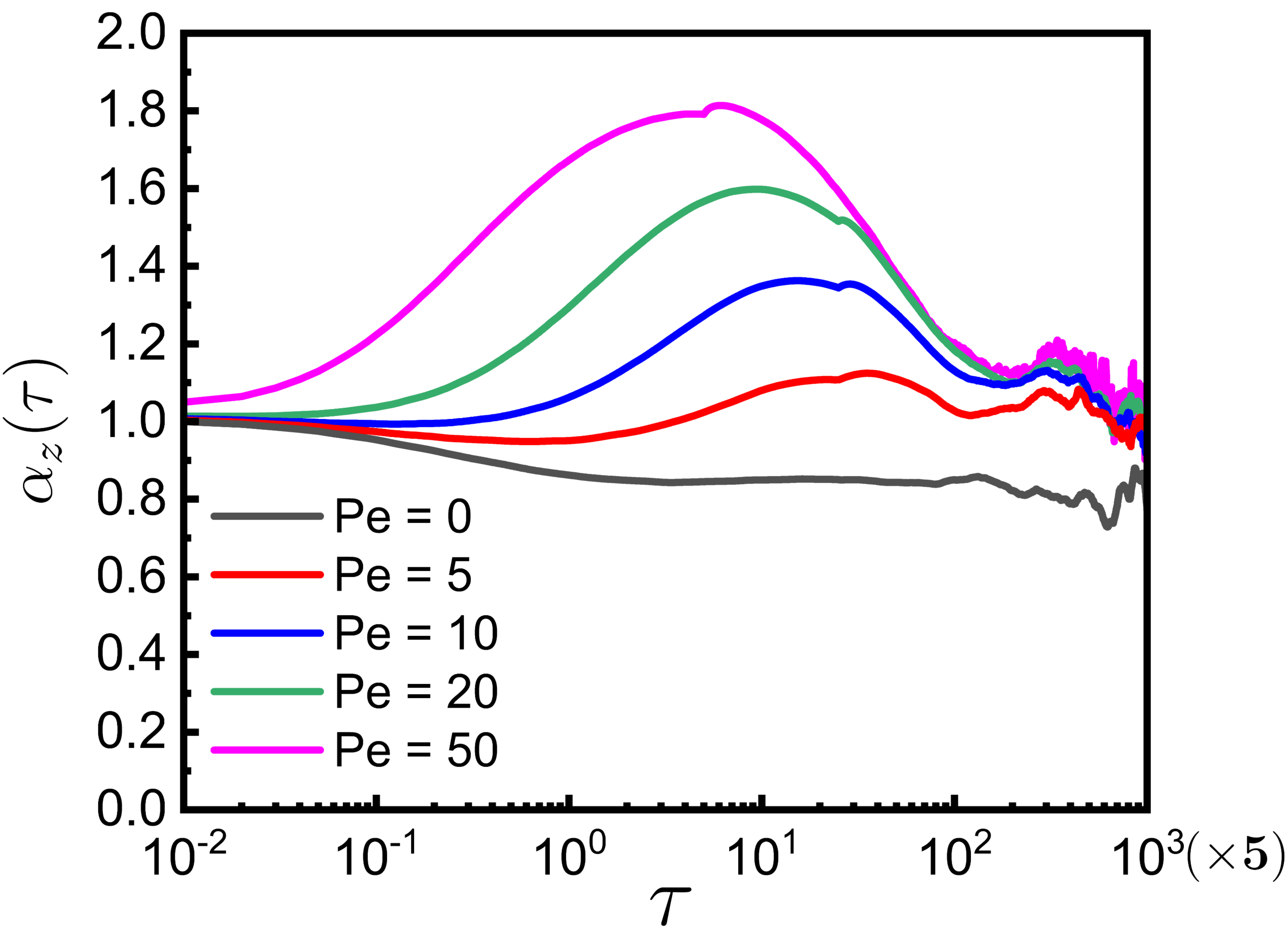} \\ 
 (a) & (b) \\
 \end{tabular}
 \caption{(a) Log–log plot of $<\delta z^2(\tau)>$ vs $\tau$ and (b) log-linear $\alpha_z(\tau)$ of the tracer particle in the polymer grafted cylindrical channel at different $\text{Pe}$ for $\epsilon = 1.5$.}
\label{fig:figure2}
\end{figure*} 

\begin{figure*}[t]
\centering
 \includegraphics[width=0.9\linewidth]{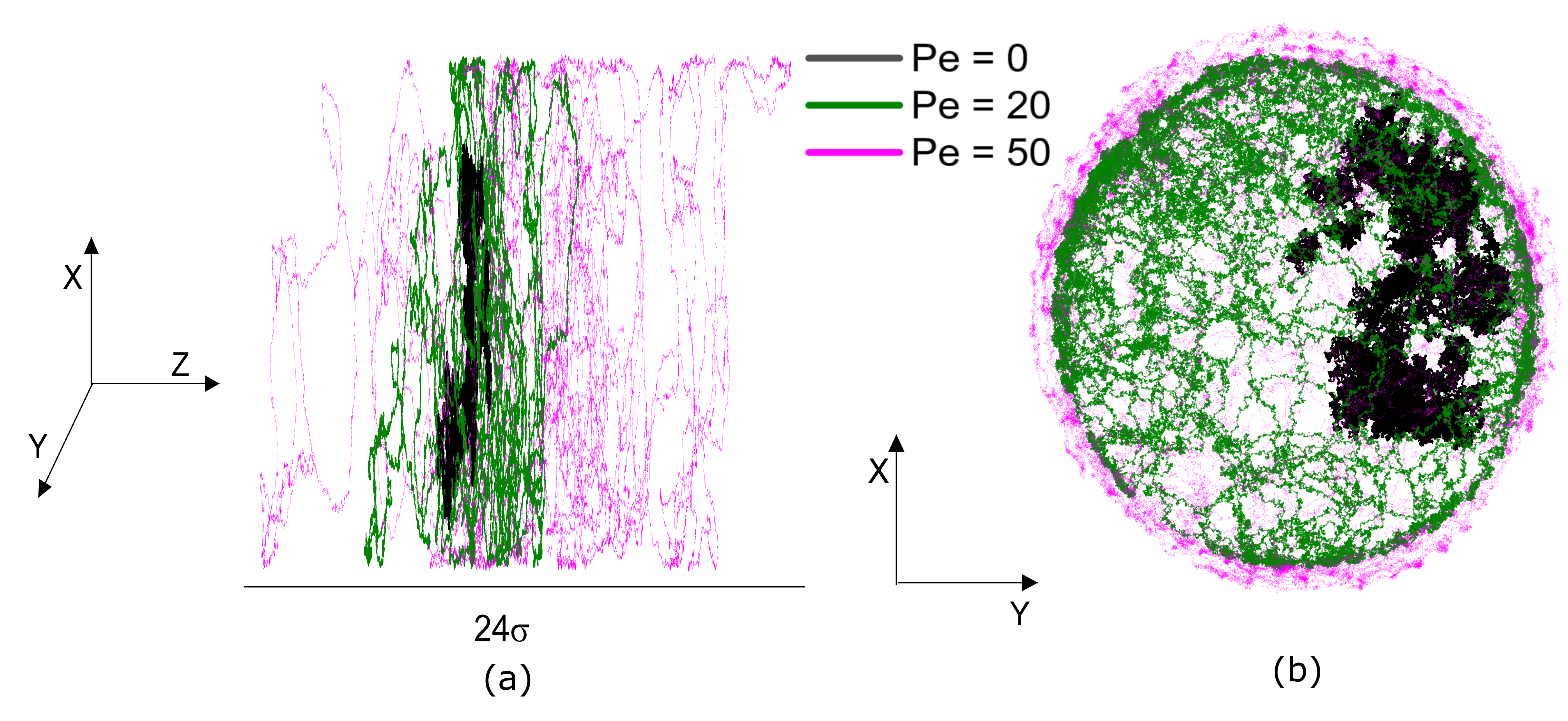} \\
 \caption{(a) Side view and (b) top view of the trajectory of the tracer particle in the polymer grafted cylindrical channel at different $\text{Pe}$.}
\label{fig:Traj1}
\end{figure*}

\noindent In order to investigate the effect of crowding and interactions, the cylindrical surface is randomly grafted with polymer chains by fixing one terminal monomer of each of the chains to one of the wall particles. Here, we study the effect of interaction strength ($\epsilon$) and the activity ($\text{Pe}$) on the dynamics of the self-propelled tracer particle. We first focus on the effect of interaction strength $\epsilon$ between the tracer particle and the polymers for the passive case, $\text{Pe} = 0$. When the polymers are purely repulsive (WCA), the tracer particle moves through the free pore-like space available between the grafted polymers along the cylinder axis causing a diffusive behavior over time with $\alpha_z(\tau) = 1$ (Fig.~\ref{fig:figure1}(a)). On the other hand, in the case when the tracer is attractive to the grafted polymers, the tracer particle goes deep into the grafted polymeric zone, leading to local trapping of the tracer (Movie1, Movie2), which slows down the dynamics and results subdiffusive ($\alpha_z(\tau) < 1$) behavior at the intermediate time (Fig.~\ref{fig:figure1}(b)). Besides this, $\left\langle\overline{\delta {(x^2+y^2)}(\tau)}\right\rangle$ also shows qualitatively similar behavior (Fig.~S2) with $\left\langle\overline{\delta z^{2}(\tau)}\right\rangle$ for the passive tracer with different $\epsilon$. The intermediate subdiffusive behavior in Fig.~\ref{fig:figure1} is more prominent upon increasing the interaction strength $\epsilon$. The cause of this intermediate subdiffusion can be interpreted as follows: the tracer particle gets trapped inside the kinks in the local configurations of the polymer at intermediate time scales. The correlated motion of the polymer is slow, and the polymer interacts strongly with the tracer leading to a slow down in tracer dynamics. When the polymer changes the configuration, the tracer escapes from these local traps. Further, to verify this, we freeze the grafted polymers and repeat the simulations. We observe that by freezing the grafted polymers, the dynamics slows down (Fig.~S3), and the subdiffusion behavior of the passive tracer is more pronounced compared to the case with mobile polymers in passive case $(\text{Pe} = 0)$. Here, the trapped tracer has to escape on its own, and the random configuration changes no longer facilitate the escape of the tracer like the case where the polymers are mobile. The tracer particle is trapped inside the frozen polymers along the radial direction and  $\left\langle\overline{\delta {(x^2+y^2)}(\tau)}\right\rangle$ gets saturated due to the confinement created by the wall (Fig.~S2(b)). On the other hand, the active tracer shows qualitatively similar behavior in mobile as well as in frozen polymers. This is caused by the tendency of the active tracer to stay close to the wall by moving through the random spaces available inside the dense polymer grafted zone to the comparatively less crowded wall region.\\

\begin{figure*}[t]
\centering
\begin{tabular}{c}
 \includegraphics[width=0.99\linewidth]{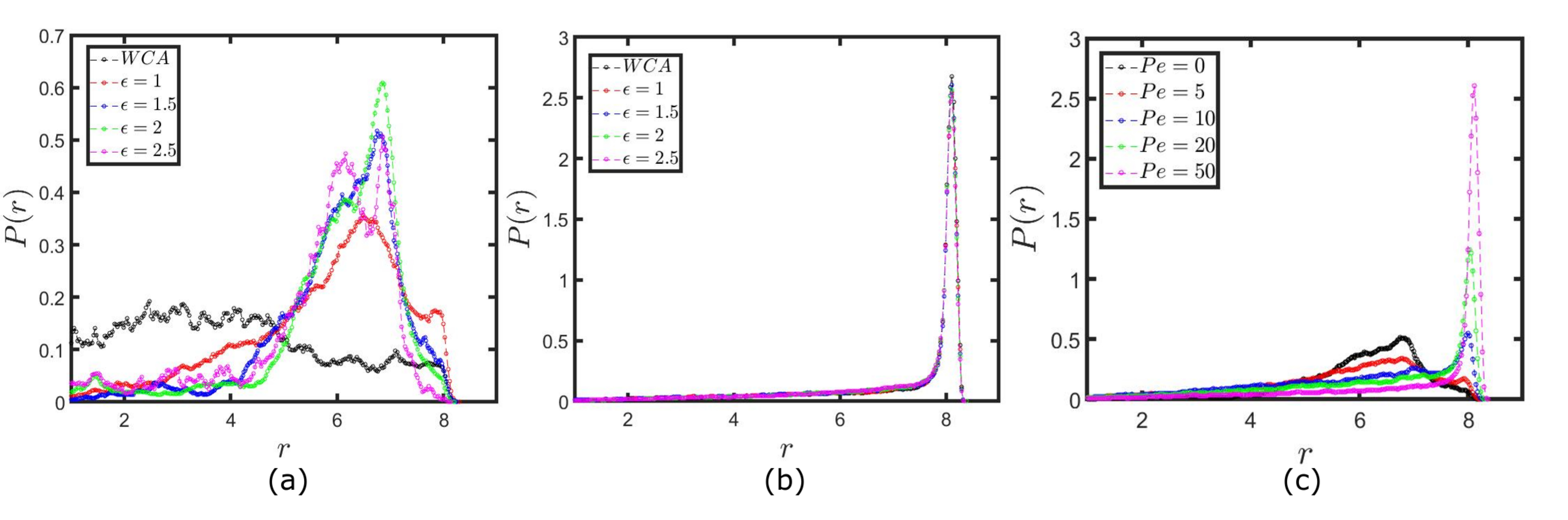} 
\end{tabular}
 \caption{$\text{P}(r)$ of the tracer particle in the polymer grafted cylindrical channel with different $\epsilon$ for (a) $\text{Pe} = 0$ and (b) $\text{Pe} = 50$ and (c) with different $\text{Pe}$ for $\epsilon = 1.5$ .}
 \label{fig:figure3}
\end{figure*}

\noindent Next, to understand the effect of activity, we take the self-propelled tracer with the interaction strength $\epsilon = 1.5$ and vary the activity ($\text{Pe}$). This is in the same spirit as in case of a self-propelled tracer (carrier) looking for a target, which has a strong affinity to the tracer. However, here we have made all the monomers of the chains equally sticky to the tracer and thus the tracer does not have a specific target zone to bind with. Here, the active tracer shows a three-step growth in $\left\langle\overline{\delta z^{2}(\tau)}\right\rangle$: short-time diffusion ($\alpha_z(\tau) = 1$) and an intermediate superdiffusion followed by a long time enhanced diffusion compared to the passive tracer ($\text{Pe} = 0$) in the polymer grafted cylindrical channel. This can be seen from Fig.~\ref{fig:figure2}. The self-propulsion turns the intermediate subdiffusion to superdiffusion. Further increasing the $\text{Pe}$, the long time value of $\left\langle\overline{\delta z^{2}(\tau)}\right\rangle$ shows a steady increase (Fig.~\ref{fig:figure2}(a)), which indicates that the activity helps the tracer to overcome the steric barrier created by the grafted polymer chains and escapes efficiently from the local traps. Interestingly, for higher $\text{Pe}$ the dynamics is much faster, and the self-propelled tracer constantly changes the directions as it more frequently encounters the polymer chains, loses the specificity, and moves deep into the grafted area near the wall of the cylinder (Movie3, Movie4) in comparison to smaller values of $\text{Pe}$, as evident from Fig.~\ref{fig:Traj1}. Thus, transport of the self-propelled particle will be faster and explore most part of the narrow channel by changing the path directions. We compute the ratio $\frac{D_{\text{Pe}}}{D_{\text{Pe} = 0}}$ to account for how much the diffusivity is enhanced due to activity. Here $D_{\text{Pe}}$ is the diffusion coefficient of the self-propelled tracer for a given $Pe$ and $D_{\text{Pe = 0}}$ is the diffusion coefficient for the passive tracer. The $\frac{D_{\text{Pe}}}{D_{\text{Pe} = 0}}$ increases with $Pe$ and shows approximately two orders of increment for $\text{Pe = 50}$ (Fig.~S4). We discover that the dynamics of the self-propelled tracer diffusing through a crowded narrow channel is governed by contrasting factors such as activity, interactions, and crowding. Crowding slows down the dynamics while the activity facilitates the tracer to overcome the local traps formed by the grafted polymers, and as a result, the transport becomes faster. However,  $\left\langle\overline{\delta {(x^2+y^2)}(\tau)}\right\rangle$ of the self-propelled tracer also exhibits a qualitatively similar pattern with the $\left\langle\overline{\delta z^{2}(\tau)}\right\rangle$ at intermediate time, but at longer time  $\left\langle\overline{\delta {(x^2+y^2)}(\tau)}\right\rangle$ saturates due to the confinement created by the wall (Fig.~S5). Apart from this, we analyze the effect of size on the dynamics by computing $\left\langle\overline{\delta z^{2}(\tau)}\right\rangle$ of the tracer particle by varying the size ($\sigma_p$) for $\text{Pe} = 20$ by keeping $\epsilon = 1.5$. Fig.~S6 clearly indicates the slowing down of the dynamics of the self-propelled tracer with its size. We also compute the velocity autocorrelation, $C_v(\tau) = \frac {\left\langle\overline{v(t+\tau).v(t)}\right\rangle}{\left\langle\overline{v^{2}(t)}\right\rangle}$, for the tracer particle at different Pe ($\epsilon = 1.5$) and stickiness (Pe = 0). We observe sharp decay in $C_v(\tau)$ with lag time $\tau$ for smaller Pe values, and shows a stronger correlation with a higher decay time with increasing Pe (Fig.~S7(a)). Besides this, the $C_v(\tau)$ falls off sharply for lower stickiness and exhibits a negative correlation at higher stickiness ($\epsilon = 2.5$) due to the trapping of the tracer particle by the grafted polymers (Fig.~S7(b))~\cite{reverey2015superdiffusion,yuan2019activity,theeyancheri2020translational}. \\

\noindent We further investigate the influence of $\epsilon$ and $\text{Pe}$ on the probability density of finding the tracer along the radial direction ($\text{P}(r)$). First, we compute $\text{P}(r)$ for different $\epsilon$ for a given $\text{Pe}$. In the case of purely repulsive tracer, $\text{P}(r)$ is maximum around the cylinder axis of the channel and gradually decreases towards the wall of the cylinder (r = 9) (Fig.~\ref{fig:figure3}(a)). Whereas, $\text{P}(r)$ changes profoundly on making the tracer attractive to the polymers. $\text{P}(r)$ has a low density around the center and has a high density at grafted polymeric region unlike the purely repulsive case (Fig.~\ref{fig:figure3}(a)). Thus, the attractive interaction significantly pushes the tracer towards the polymer grafted region from the central region of the cylinder and is reflected in Fig.~\ref{fig:figure3}(a) by the distribution peaks shifting to larger values of $r$ with increasing values of $\epsilon$. This indicates that the repulsive tracer spends more time at the pore-like space in the central region of the channel and the probability decreases along the radial direction, while the attractive tracer prefers to stay close to the wall of the polymer grafted cylinder. Next, we vary $\epsilon$ by keeping $\text{Pe} = 50$ for the self-propelled tracer particle. $\text{P}(r)$ shows the density profile with sharp peaks located close to the wall of the cylinder ($r = 8 \sigma$) for $\text{Pe} = 50$ irrespective of $\epsilon$ as shown in Fig.~\ref{fig:figure3}(b). This implies that it is equally probable to have the tracer near the wall of the cylindrical channel for all values of $\epsilon$ for larger $\text{Pe}$. It is consistent with the trajectory of the tracer in Fig.~\ref{fig:figure3}(b), which shows repeatedly changing paths inside the channel for increasing value of $\text{Pe}$. Next, we vary the activity of the tracer for a fixed value of $\epsilon$. A more broader and flat density profile with increasing $\text{Pe}$ for the self-propelled tracer as shown in Fig.~\ref{fig:figure3}(c), while for very high $\text{Pe}$ the active tracer prefer to stay close to the wall which results into a narrow distribution. It is equally probable to have the tracer everywhere inside the cylindrical channel for higher values of $\text{Pe}$. \\
\begin{figure*}[t]
\centering
\begin{tabular}{cc}
 \includegraphics[width=0.44\linewidth]{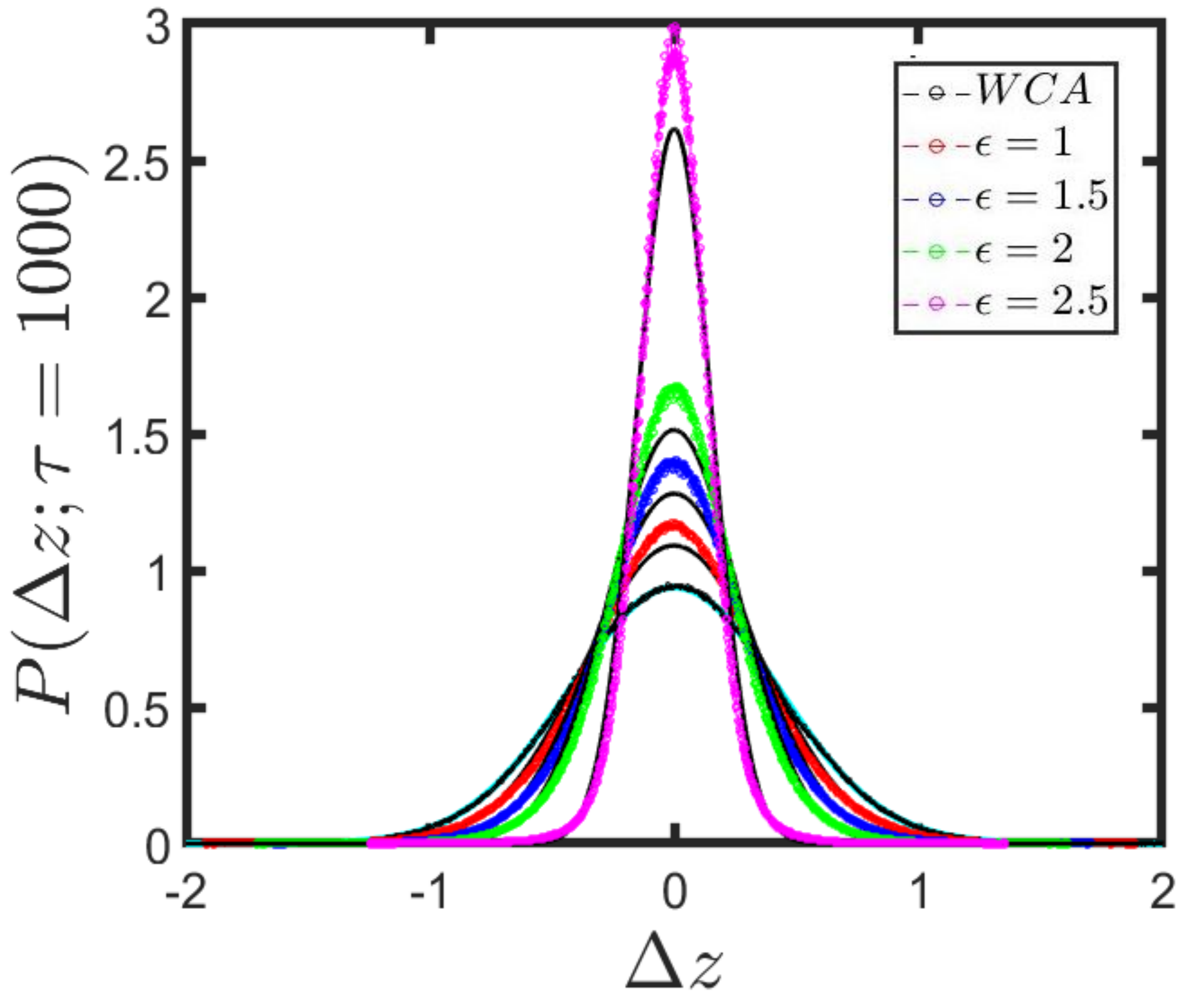} & \includegraphics[width=0.44\linewidth]{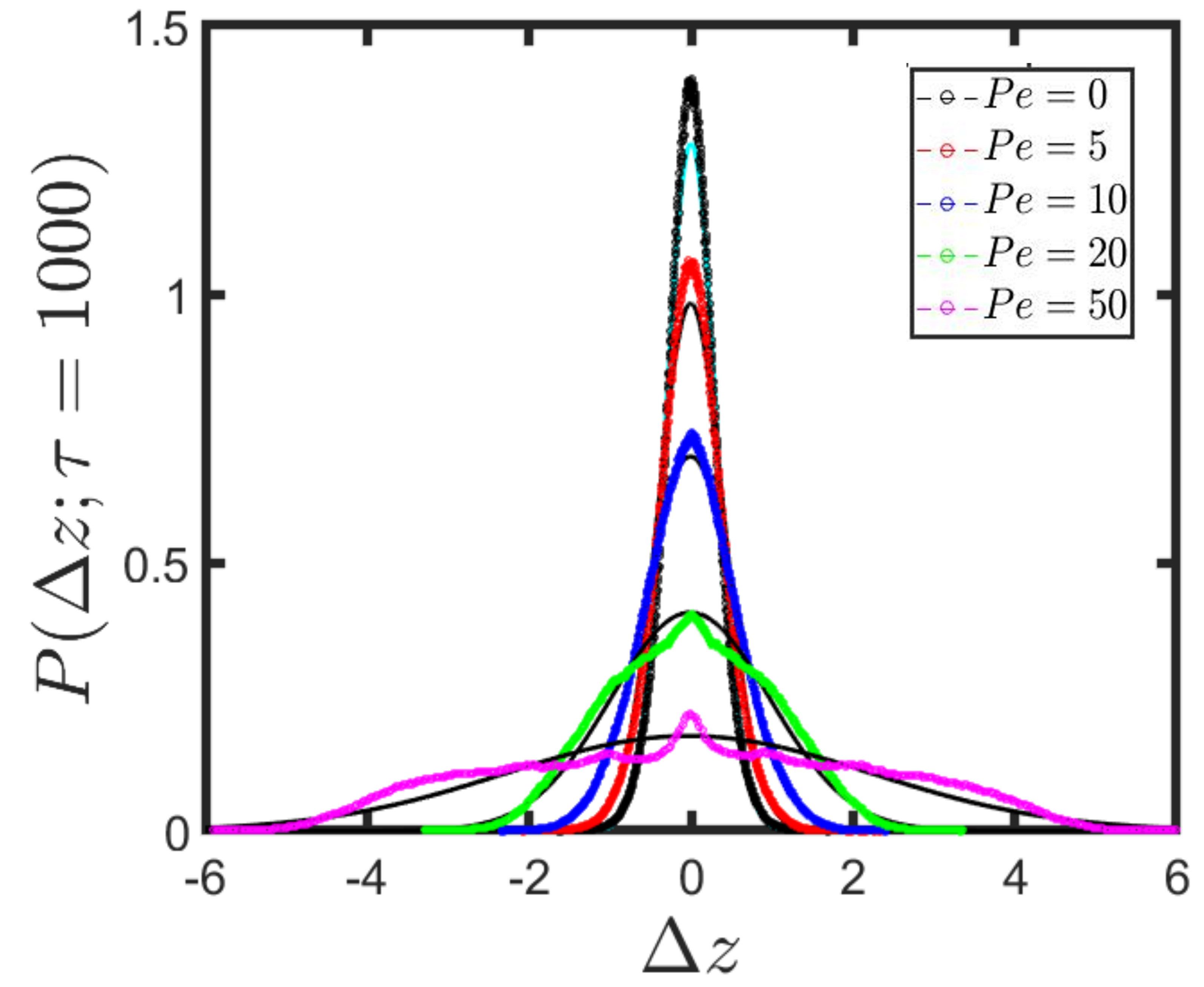} \\ 
 (a) & (b) \\
\end{tabular}
 \caption{$\text{P}(\Delta z; \tau)$ of the tracer particle in the polymer grafted cylindrical channel with (a) different $\epsilon$ for $\text{Pe} = 0$ and (b) for different $\text{Pe}$ with $\epsilon = 1.5$. The solid lines (black and cyan) represent the Gaussian fittings.}
 \label{fig:figure4}
\end{figure*} 

\noindent Subsequently, to achieve a deeper understanding of the underlying dynamics, we compute the probability distribution function $\text{P}(\Delta z; \tau) \equiv \left\langle \delta(\Delta z - (z(t+\tau) - z(t)))\right\rangle$ of the tracer particle’s displacement in one dimension (along z-direction) ~\cite{kumar2019transport}, where $\text z(t + \tau)$ and z(t) are the positions of the tracer along the cylinder axis at a time $( t + \tau)$ and t, respectively. $\text P(\Delta z;t)$ corresponds to the time and ensemble-averaged self-part of the van-Hove function. In Fig.~\ref{fig:figure4} we plot $\text P(\Delta z;t)$ with the corresponding Gaussian distribution functions for the free Brownian motion, \text{$P(\Delta z) = \frac {1}{\sqrt {2\pi \langle \Delta z^2 \rangle}} \text {exp} (-\frac {\Delta z^2}{2 \langle \Delta z^2 \rangle })$}. As a first step, we analyze $\text P(\Delta z;t)$ of the passive tracer ($\text{Pe} = 0$) for different $\epsilon$. Fig.~\ref{fig:figure4}(a) depicts that due to the confined motion of the tracer, $\text P(\Delta z;t)$ becomes narrower with increasing $\epsilon$. This is more prominent for higher values of $\epsilon$. We show that the profiles can be well fitted by Gaussian distributions (solid lines in Fig.~\ref{fig:figure4}(a)). Moreover, we investigate whether the effect of activity will reshape the probability density profile of the tracer displacement by evaluating $\text P(\Delta z;t)$ for the self-propelled particle by varying the $\text{Pe}$ for a constant $\epsilon$. As compared to the passive case (${Pe} = 0$), the width of the $\text P(\Delta z;t)$ is evidently wider, indicating the activity-induced escaping of the self-propelled tracer and enhanced diffusivity. The distribution becomes remarkably broader as a function of $\text{Pe}$ (Fig.~\ref{fig:figure4}(b). More importantly, the motion of the self-propelled tracer becomes non-Gaussian at higher $\text{Pe}$ owing to the self-propelled motion of the tracer. The deviation from the Gaussianity is more pronounced for larger $\text{Pe}$ values. A similar behavior is observed for the displacement along radial direction shown in Fig.~S8.

\section{Conclusions}
\noindent In this work, we have extensively analyzed the dynamics of a self-propelled as well as a passive tracer in a cylindrical channel grafted with polymers by varying the interaction strength between the tracer and the polymers and the activity of the tracer. Our simulation results show that the passive tracer exhibits an intermediate subdiffusion when the interaction is attractive, unlike the purely repulsive case where the tracer is diffusive at all times. The intermediate subdiffusion gets more pronounced upon increasing the strength of the attractive interaction. The tracer particle is trapped inside the local minima of the rugged energy landscape created by the local configurations of the polymers cause this intermediate subdiffusive behavior~\cite{samanta2016tracer}.  In general, the trapped tracer particle follows the motion of the chain leading to a slow down of dynamics. The change in configurations of polymers facilitates the escape events of tracer from the transient traps and again shows the diffusive character. Subsequently, this is justified by the long-lived subdiffusive motion observed when we freeze the grafted polymer, where the polymers no longer change their configurations. More importantly, our simulations reveal that there is an enhancement in diffusivity on changing the passive tracer to a self-propelled one, in the presence of grafted polymers. The diffusivity increases as a function of the activity. A clear difference in dynamics is observed for the self-propelled tracer compared to the passive tracer. Here the activity turns the intermediate subdiffusion to superdiffusion. This superdiffusion is more notable for higher $\text{Pe}$, and the dynamics is faster even though the self-propelled tracer constantly changes the directions inside the cylindrical channel. This observation is further supported by the probability density to find the tracer along the radial direction, which shows increasingly broader distributions with the increase in activity or the interaction strength. The tracer moves deep into the grafted polymeric zone, which leads to a strong intermediate subdiffusion with increasing interaction, while the increase of activity also leads to broader distributions but faster dynamics and exhibits a superdiffusion at intermediate time. The dynamics of the self-propelled tracer displays deviation from the Gaussianity at the higher range of activities, whereas the passive tracer dynamics with different interaction strength remains as Gaussian. Our study discloses that the balance between the activity and interactions facilitates the transport of tracer through the crowded narrow channel.\\

\noindent In a nutshell, our present work focuses on the anomalous diffusive dynamics of a self-propelled particle through a crowded channel. Our findings can provide useful insights into the active transport facilitated by biological pumps~\cite{drory2006emerging} or motor molecules~\cite{soppina2009tug,alberts1998cell,kinbara2005toward}. Drawing motivation from the performance of biological motors and pumps, researchers have designed hybrid bio-synthetic nano or micro transporters by incorporating motor proteins into artificial systems~\cite{hess2001molecular,loget2010propulsion,van2007motor,hiratsuka2006microrotary}. Previous studies revealed that a unidirectional transport of materials occurs when the cargo is attached to kinesin~\cite{limberis2001polarized,bohm2001motor}. Thus, artificial self-powered nano or microdevices have potential applications in targeted delivery and are used as a carrier for cargo transportation and delivery purposes~\cite{patra2013intelligent}. So they are subjected to a range of topological constraints and interactions while performing the assigned tasks. It is highly demanding to have a better understanding and tuning of highly selective transport of macromolecules through cellular channels such as the NPC, mucous membranes, and the extracellular matrix that depends on speed, size, and binding affinity. We believe that the main features of our findings stay valid qualitatively for understanding the active transport through crowded channels. More work in the future will be required to manifest the hidden mechanisms underlying the transport using active agents. Further studies of active transport through crowded narrow channels by incorporating specific and non-specific binding zones on polymers are anticipated in future.

\begin{acknowledgements}
\noindent  R.S. thanks CSIR for a fellowship. L.T. thanks UGC for a fellowship. R.C. acknowledge SERB, India, $via$ Project No. MTR/2020/000230 under MATRICS scheme and IRCC-IIT Bombay (Project No. RD/0518-IRCCAW0-001) for funding. R.S. acknowledges Praveen Kumar for the helpful discussions.  We acknowledge the SpaceTime-2 supercomputing facility at IIT Bombay for the computing time.  \\
\end{acknowledgements}

\noindent \textbf{DATA AVAILABILITY} \\

\noindent The data that supports the findings of this study are available within the article [and its supplementary material]. \\

\renewcommand{\thefigure}{S\arabic{figure}}
\setcounter{figure}{0}
\appendix
\noindent \textbf{SUPPLEMENTARY MATERIAL} \\

\noindent \textbf{Figures}\\

\begin{figure*}[h]
\centering
 %\begin{tabular}{cc}
  \includegraphics[width=0.5\linewidth]{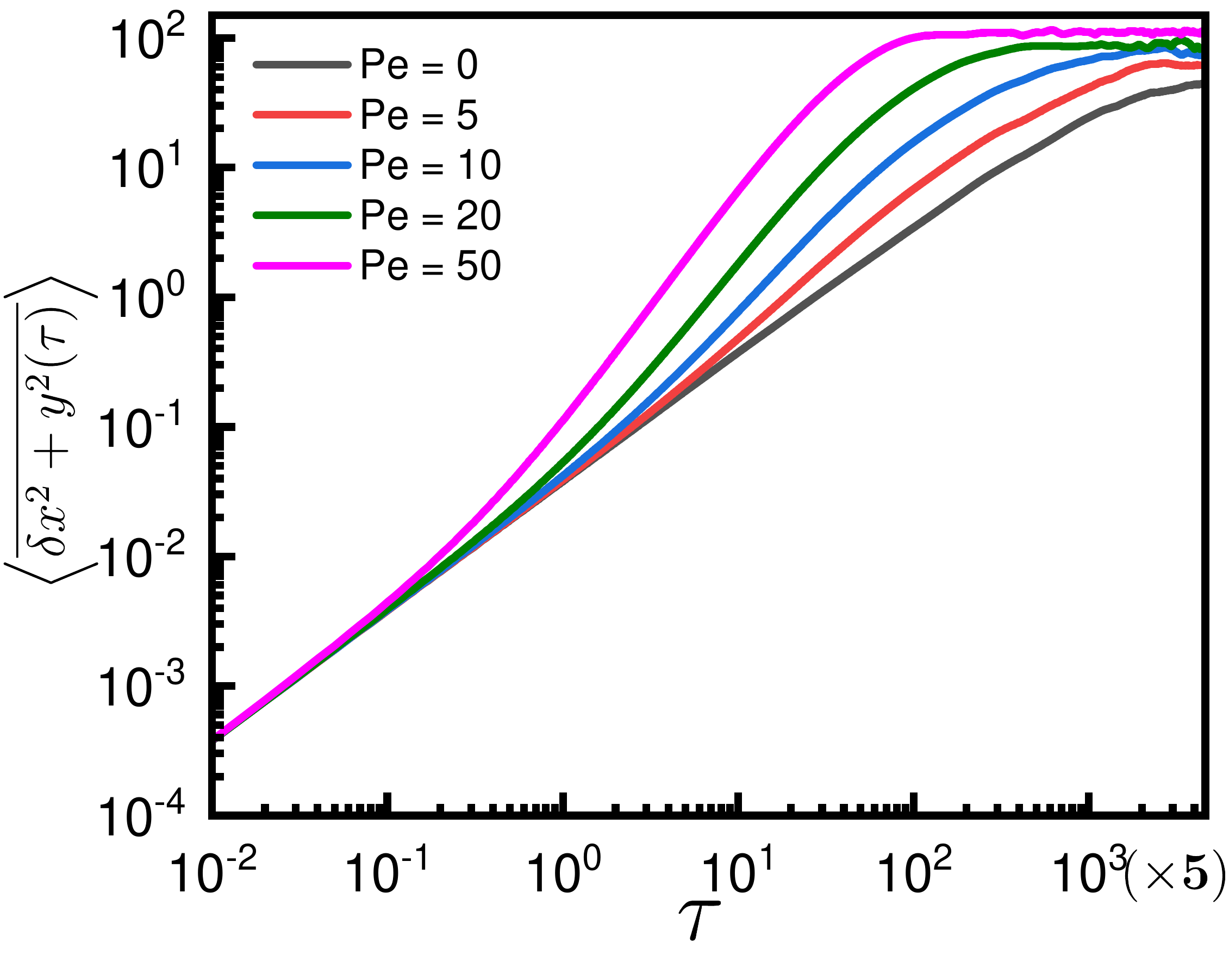}
 % \includegraphics[height=6cm]{Figure_E2.5_Vert.pdf} \\ [1ex]
  %(a) & (b)
 % \end{tabular}
   \caption{Log–log plot of $\left\langle\overline{\delta {(x^2+y^2)}(\tau)}\right\rangle$ vs $\tau$ of the tracer particle in cylindrical channel without polymers, at different $\text{Pe}$.}
  \label{fig:Figure1}
\end{figure*}

\begin{figure*}[h]
\centering
 %\begin{tabular}{cc}
  \includegraphics[width=0.5\linewidth]{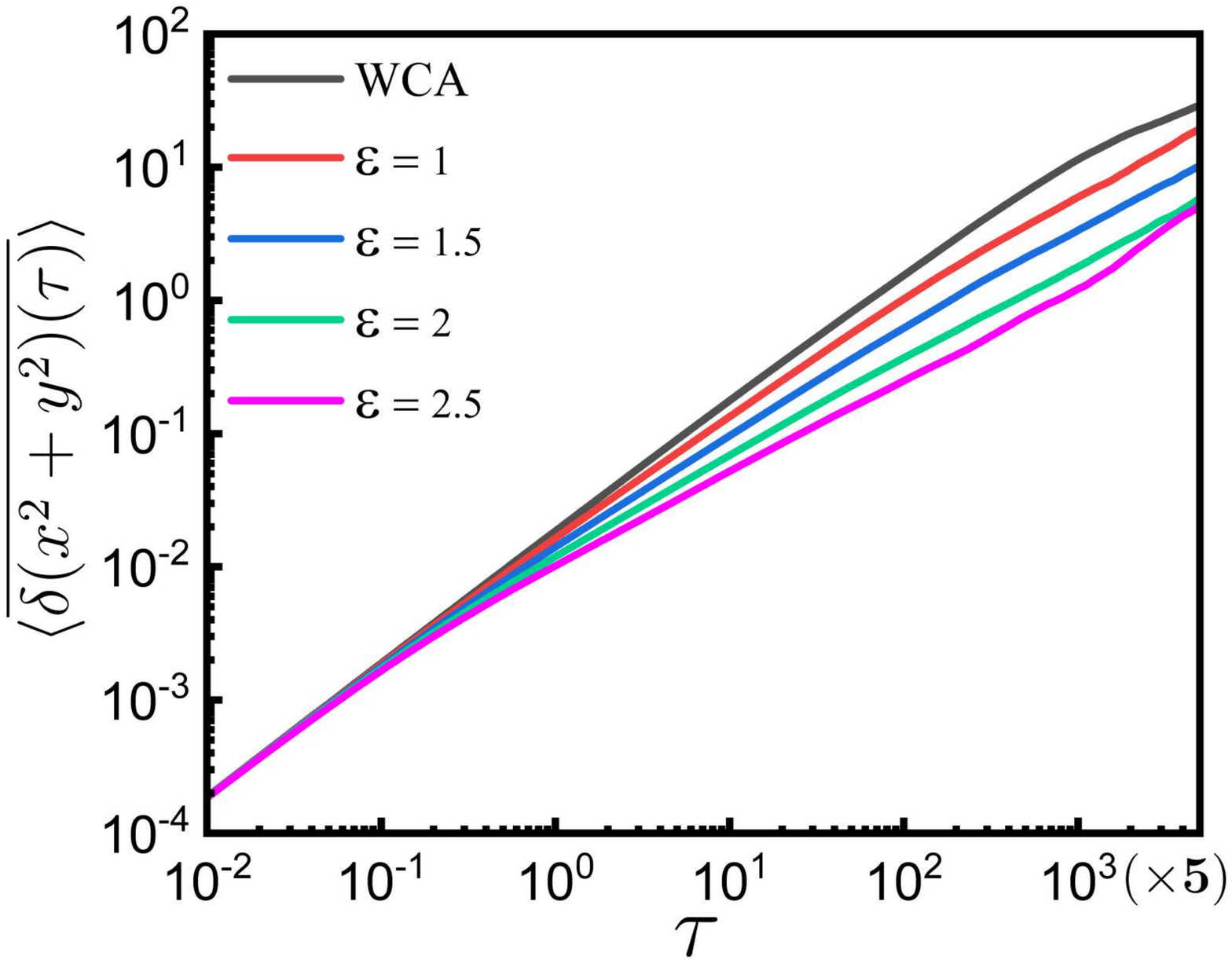}
 % \includegraphics[height=6cm]{Figure_E2.5_Vert.jpg} \\ [1ex]
  %(a) & (b)
 % \end{tabular}
   \caption{Log–log plot of $\left\langle\overline{\delta {(x^2+y^2)}(\tau)}\right\rangle$ vs $\tau$ of the passive tracer particle in the polymer grafted cylindrical channel with different $\epsilon$.}
  \label{fig:Figure2}
\end{figure*}

\begin{figure*}[t]
\centering
\begin{tabular}{cc}
 \includegraphics[width=0.5\linewidth]{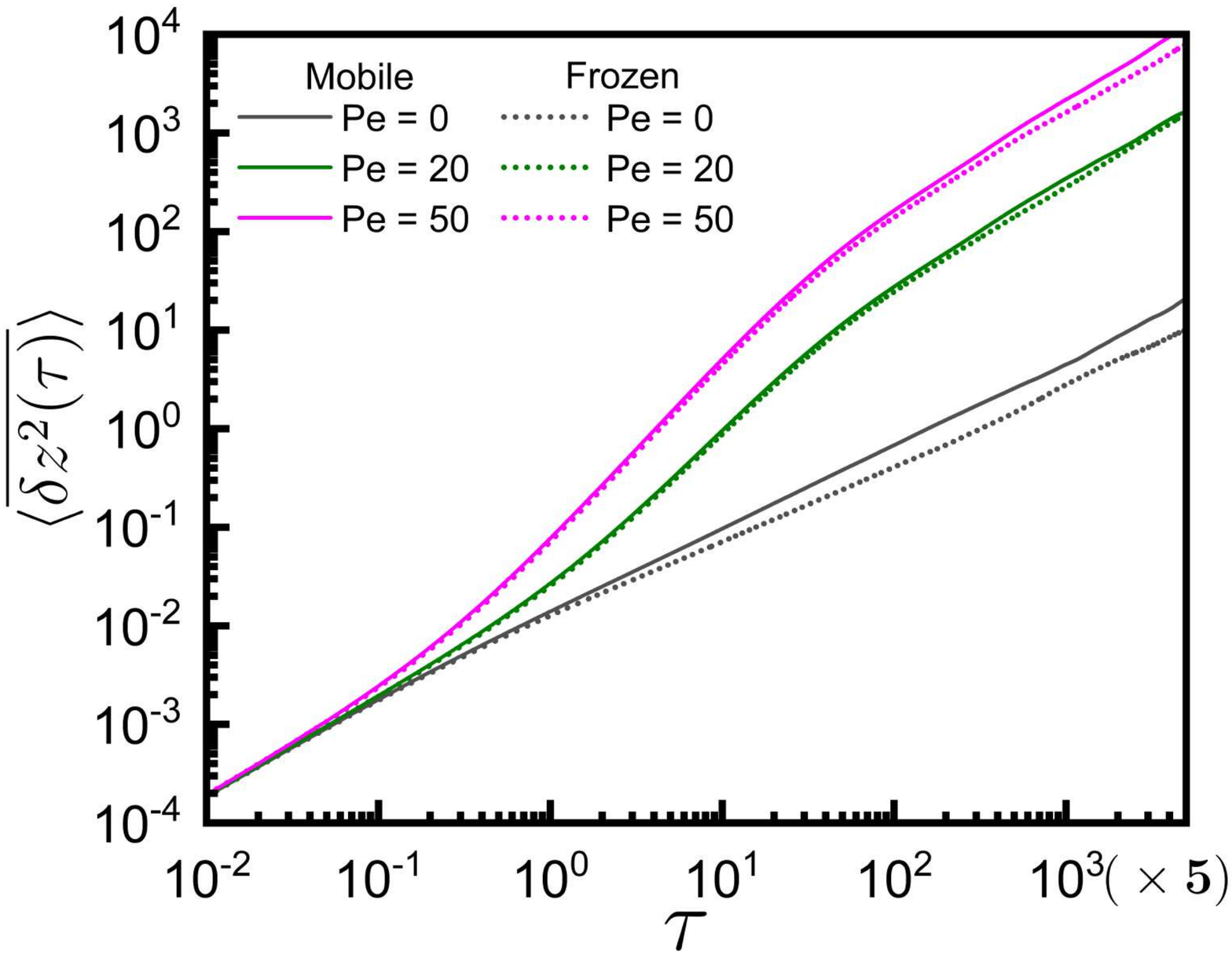} & \includegraphics[width=0.52\linewidth]{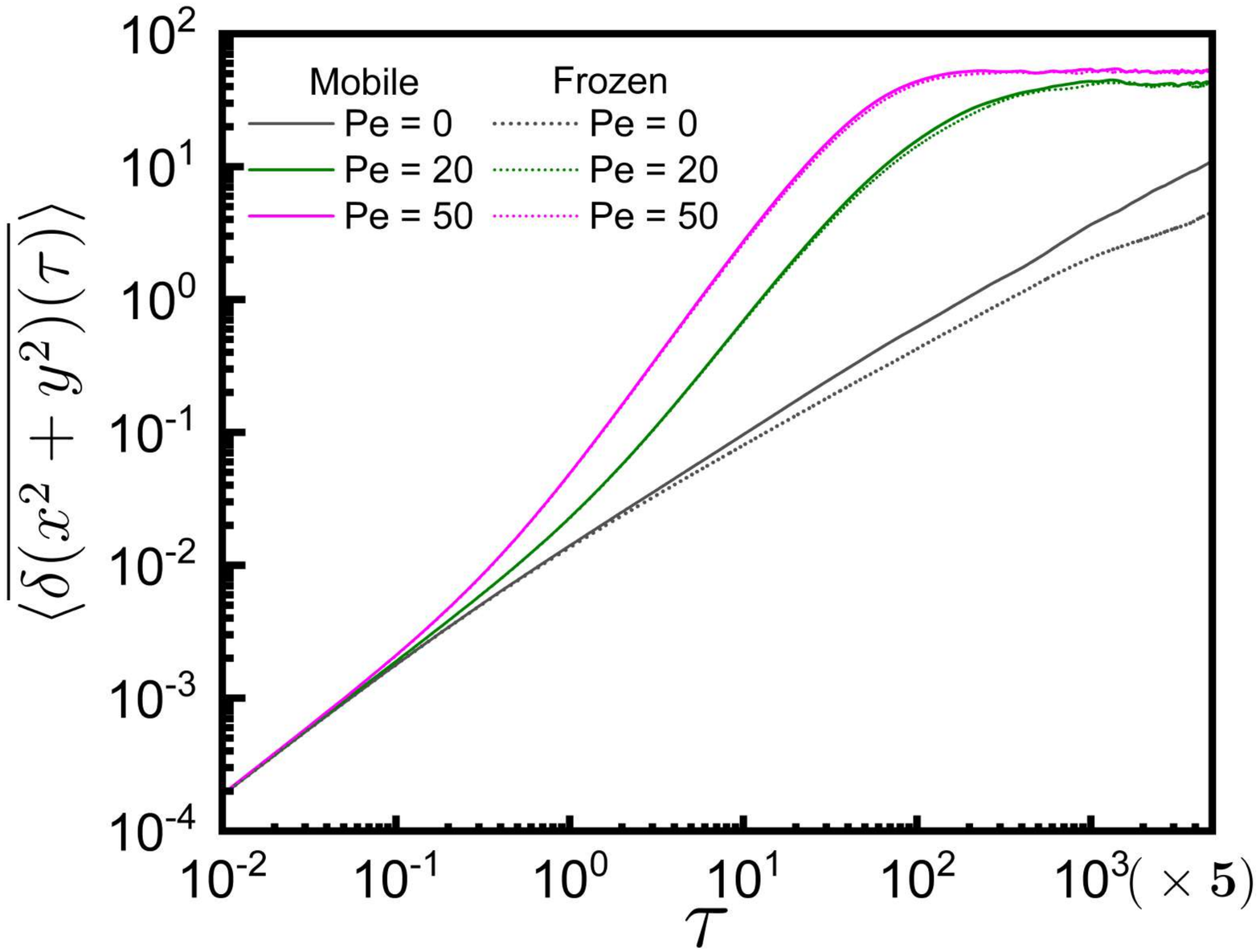} \\ 
 (a) & (b) \\
\end{tabular}
 \caption{Log–log plot of (a) $<\delta z^2(\tau)>$ vs $\tau$ and (b) $\left\langle\overline{\delta {(x^2+y^2)}(\tau)}\right\rangle$ vs $\tau$ for the tracer particle in the frozen polymer (dotted lines) grafted cylindrical channel and in the mobile polymer (solid lines) grafted cylindrical channel at different $\text{Pe}$ for $\epsilon = 1.5$.}
 \label{fgr:figure2}
\end{figure*}

\begin{figure*}[t]
\centering
 %\begin{tabular}{cc}
  \includegraphics[width=0.5\linewidth]{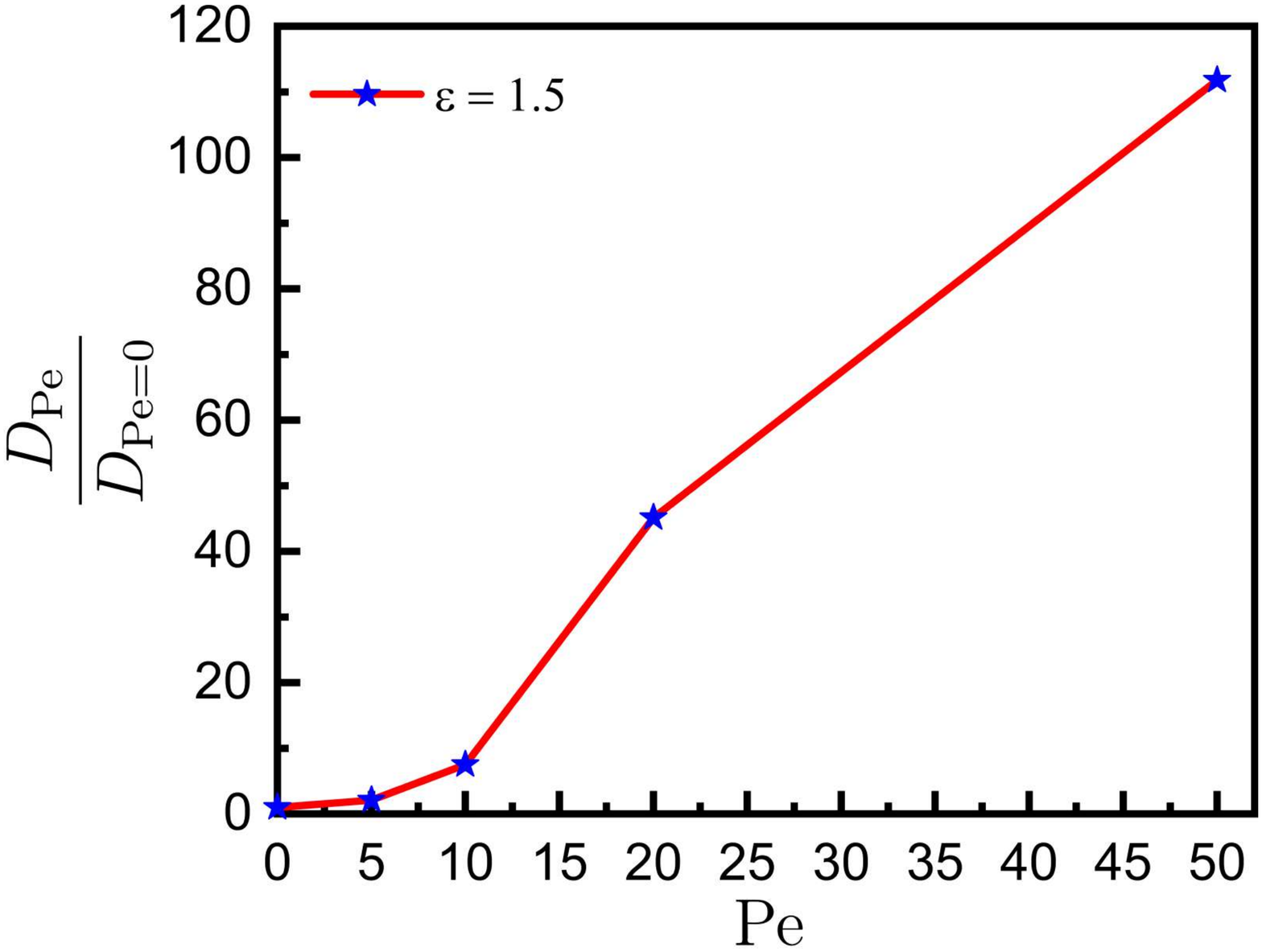}
 % \includegraphics[height=6cm]{Figure_E2.5_Vert.jpg} \\ [1ex]
  %(a) & (b)
 % \end{tabular}
  \caption{ $\frac{D_{\text{Pe}}}{D_{\text{Pe} = 0}}$ of the tracer particle in the polymer grafted cylindrical channel at different $\text{Pe}$ for $\epsilon = 1.5$.}
  \label{fig:Figure3}
\end{figure*}

\begin{figure*}[t]
\centering
 %\begin{tabular}{cc}
  \includegraphics[width=0.5\linewidth]{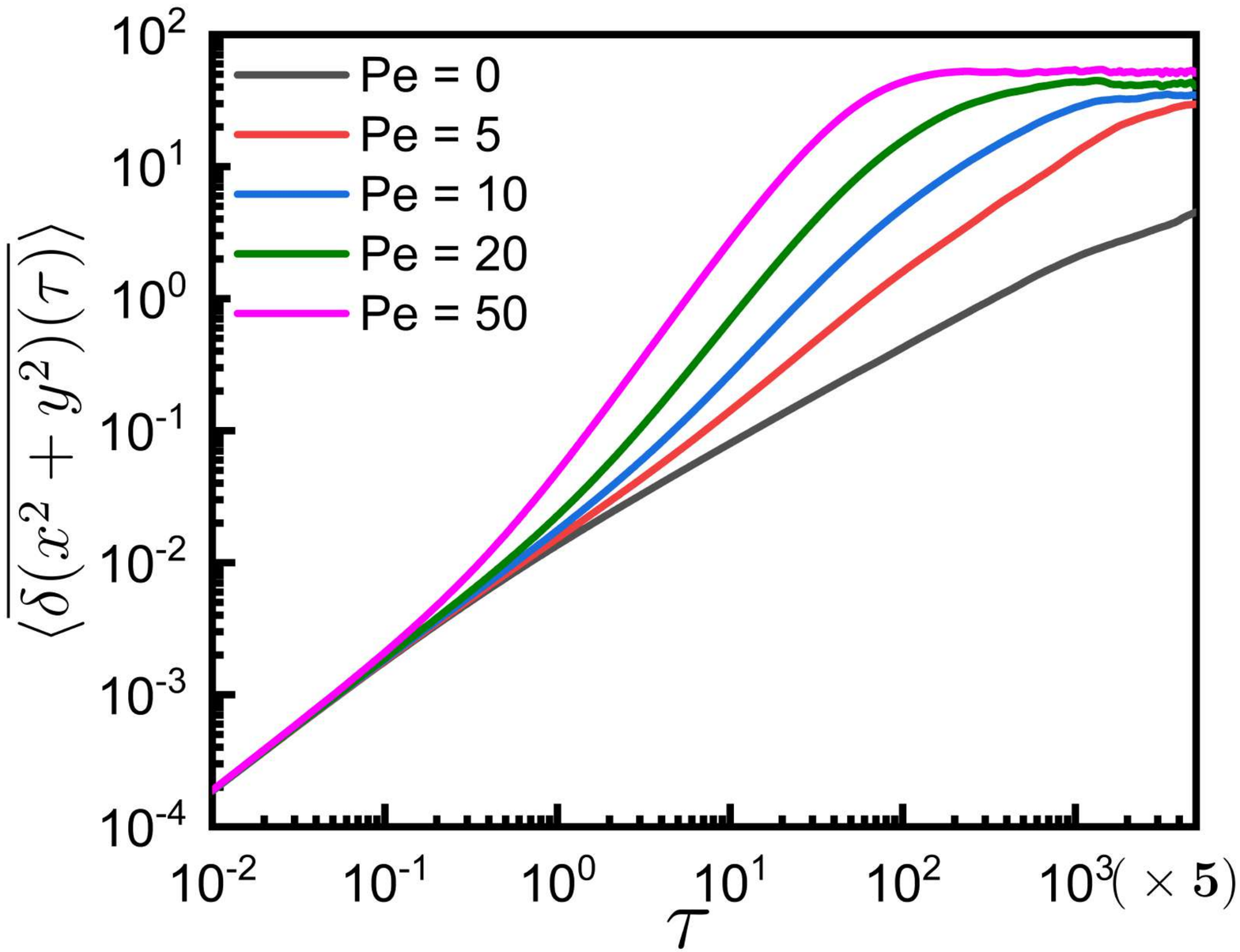}
 % \includegraphics[height=6cm]{Figure_E2.5_Vert.jpg} \\ [1ex]
  %(a) & (b)
 % \end{tabular}
  \caption{ Log–log plot of $\left\langle\overline{\delta {(x^2+y^2)}(\tau)}\right\rangle$ vs $\tau$ of the tracer particle in the polymer grafted cylindrical channel at different $\text{Pe}$ for $\epsilon = 1.5$.}
  \label{fig:Figure4}
\end{figure*}

\begin{figure*}[t]
\centering
\begin{tabular}{cc}
 \includegraphics[width=0.5\linewidth]{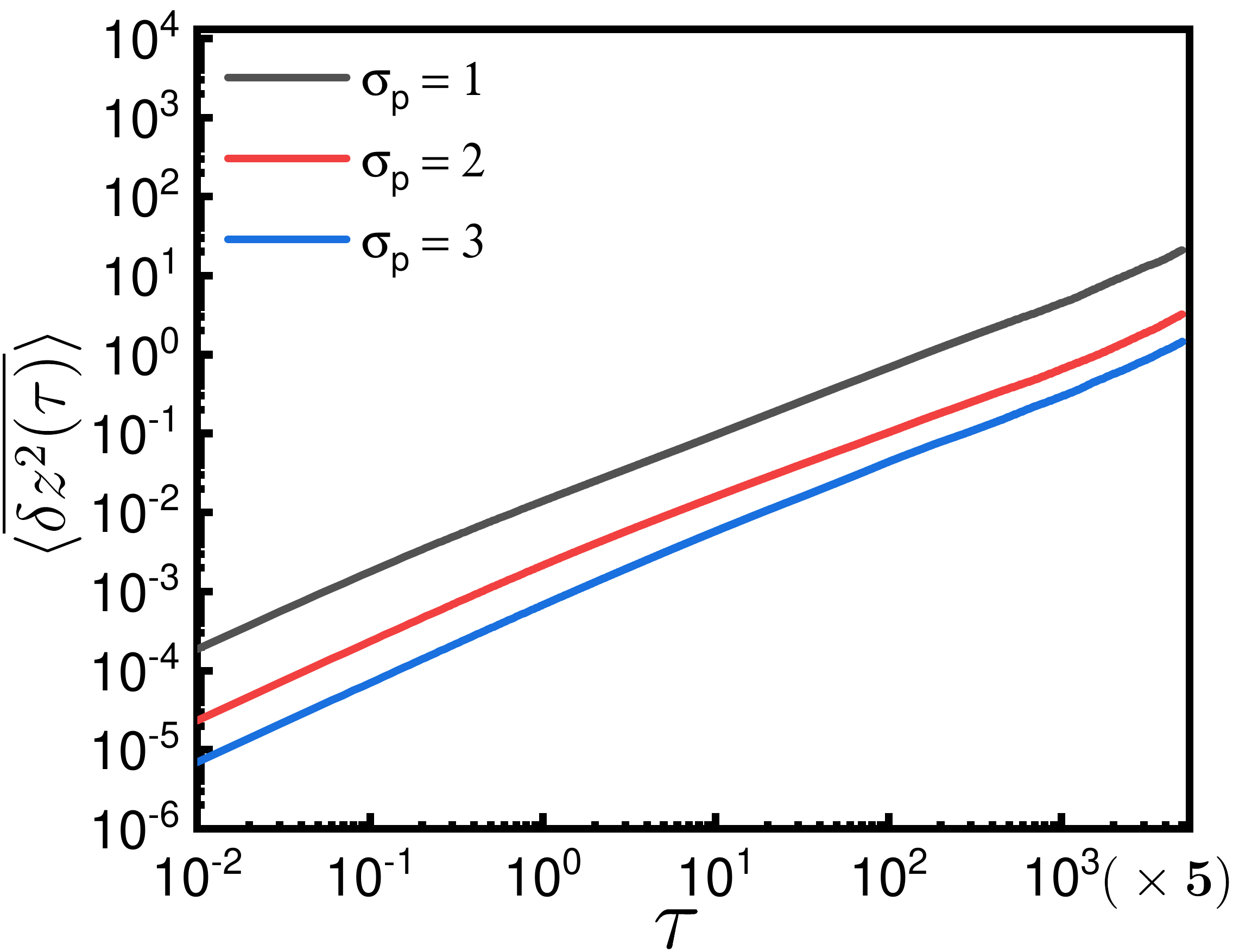} & \includegraphics[width=0.51\linewidth]{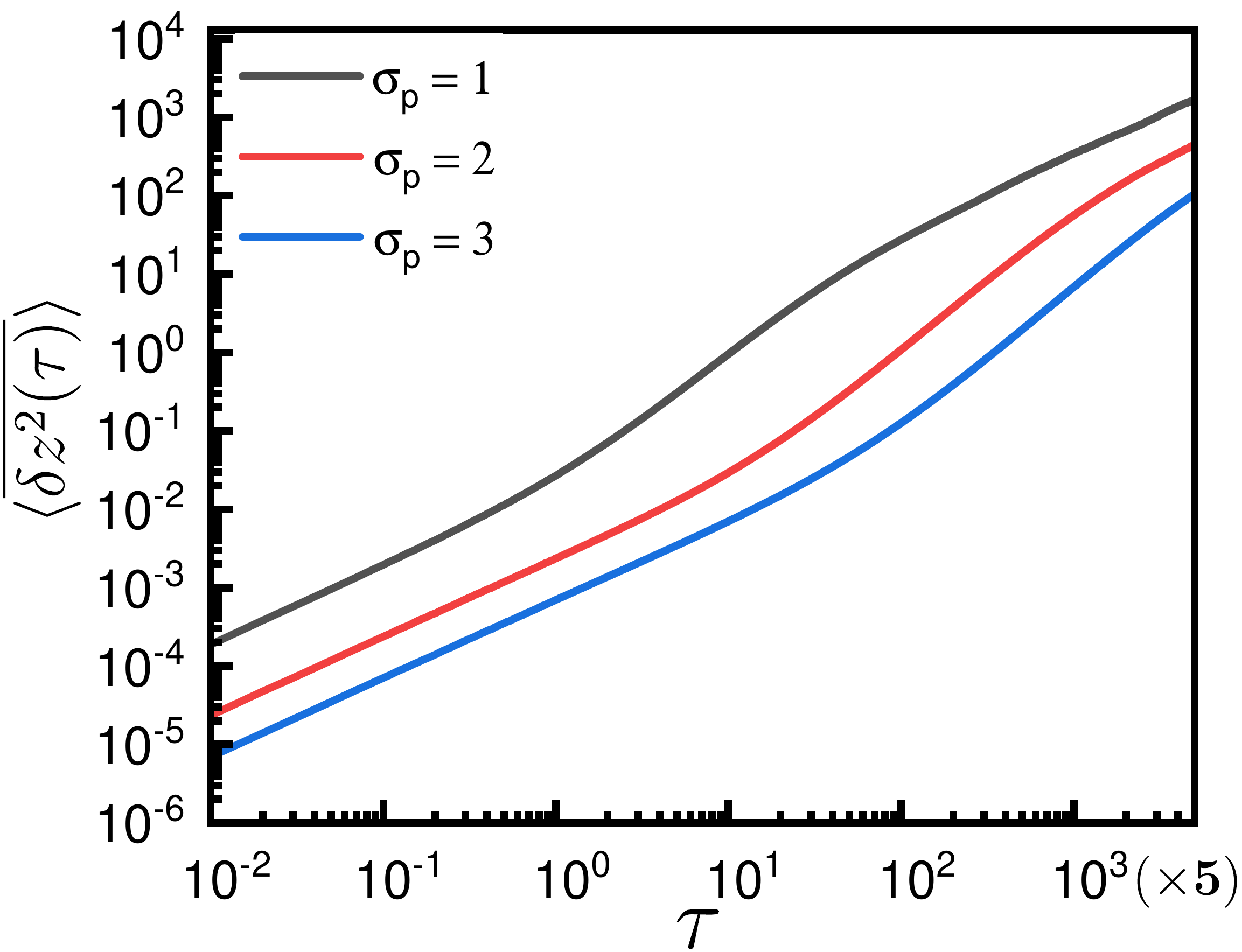} \\ 
 (a) & (b) \\
\end{tabular}
 \caption{Log–log plot of $<\delta z^2(\tau)>$ vs $\tau$ of the tracer particle with different size for (a) $\text{Pe} = 0$ and (b) $\text{Pe} = 20$ with $\epsilon = 1.5$.}
 \label{fgr:figure5}
\end{figure*}

\begin{figure*}[t]
\centering
\begin{tabular}{cc}
 \includegraphics[width=0.5\linewidth]{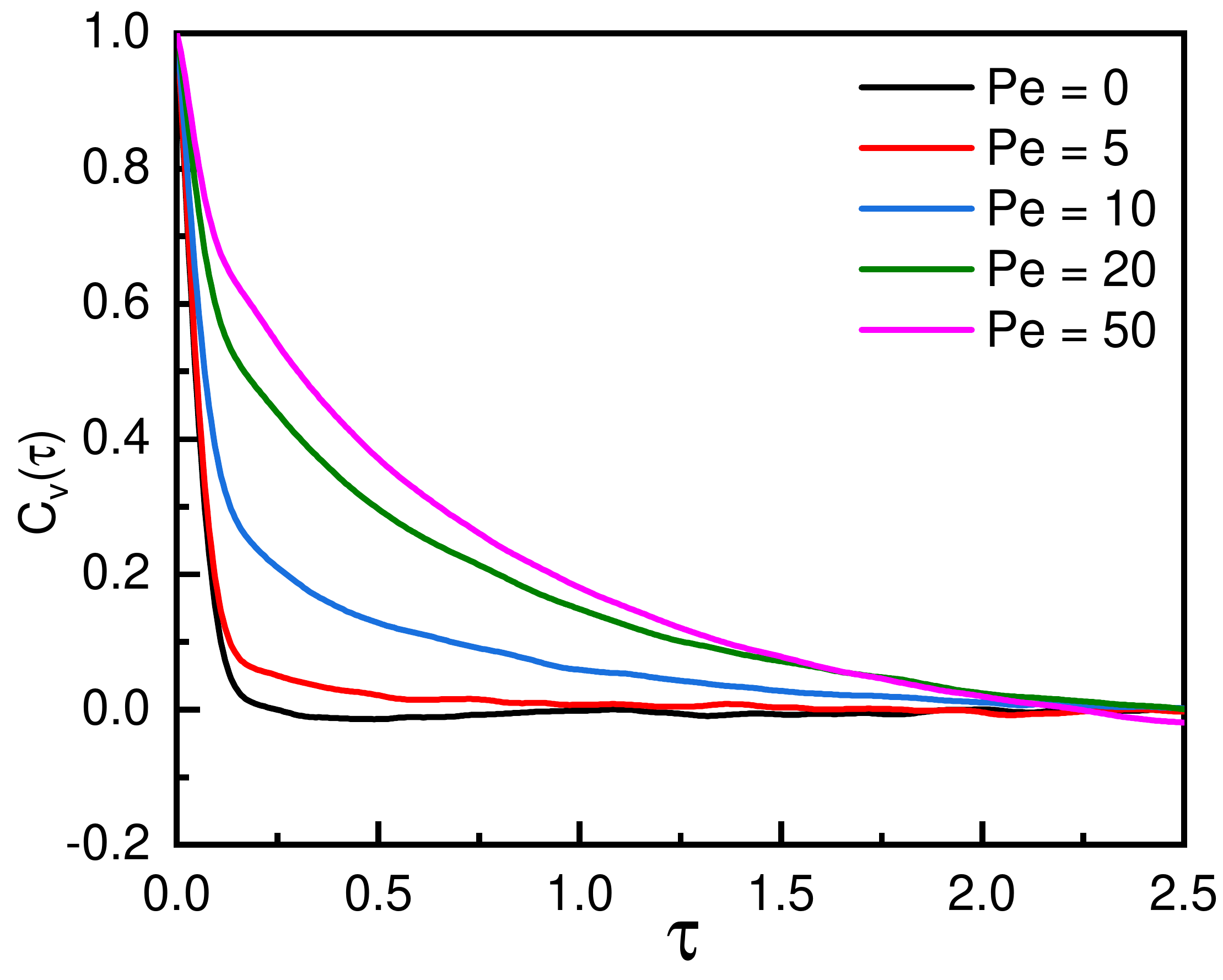} & \includegraphics[width=0.51\linewidth]{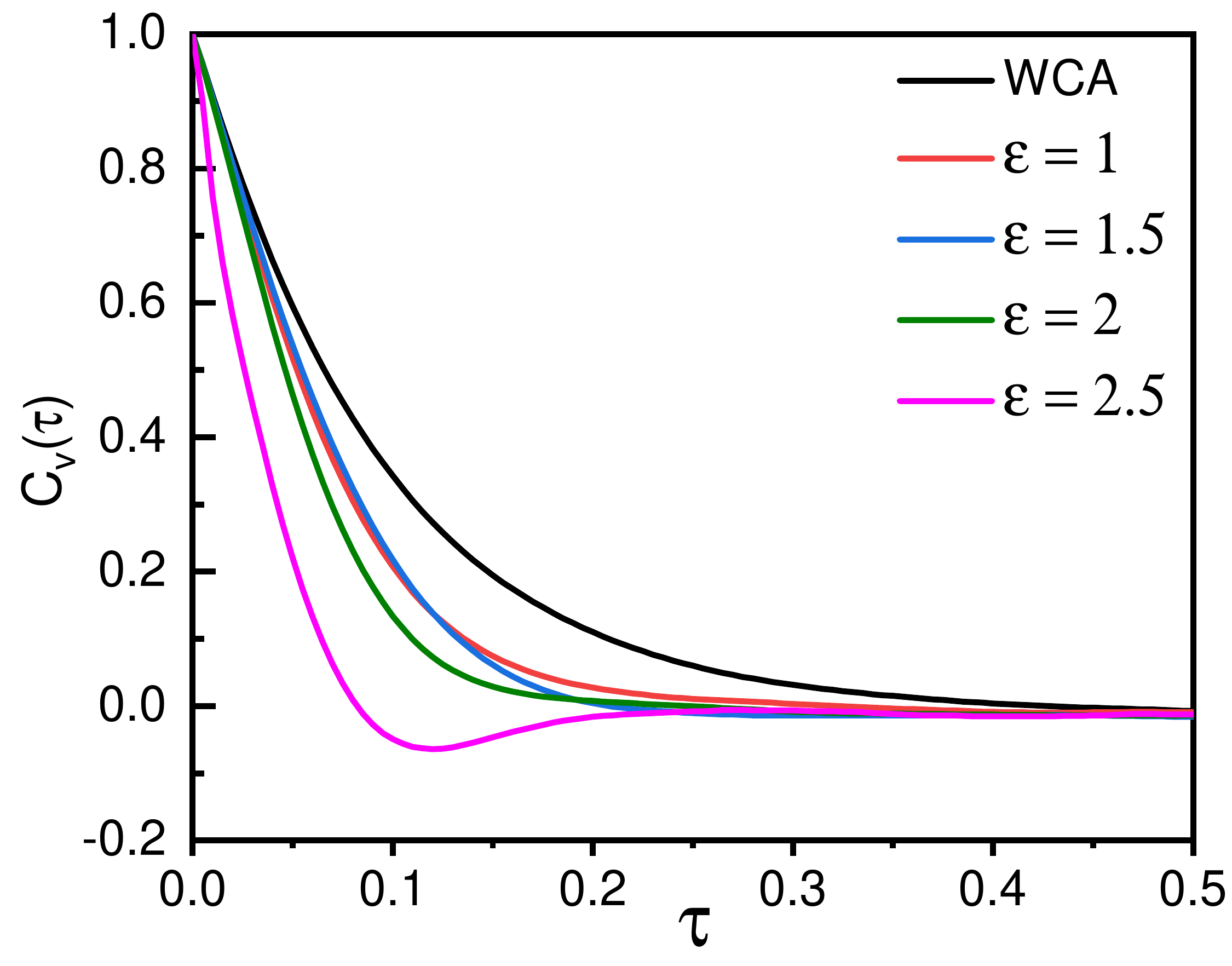} \\ 
 (a) & (b) \\
\end{tabular}
 \caption{$C_v(\tau)$ $vs$ lag time $(\tau)$ of (a) active tracer at different activity for $\epsilon = 1.5$ and (b) passive tracer with different stickiness in low friction limit.}
 \label{fgr:figure6}
\end{figure*}

\begin{figure*}[t]
\centering
\begin{tabular}{cc}
 \includegraphics[width=0.48\linewidth]{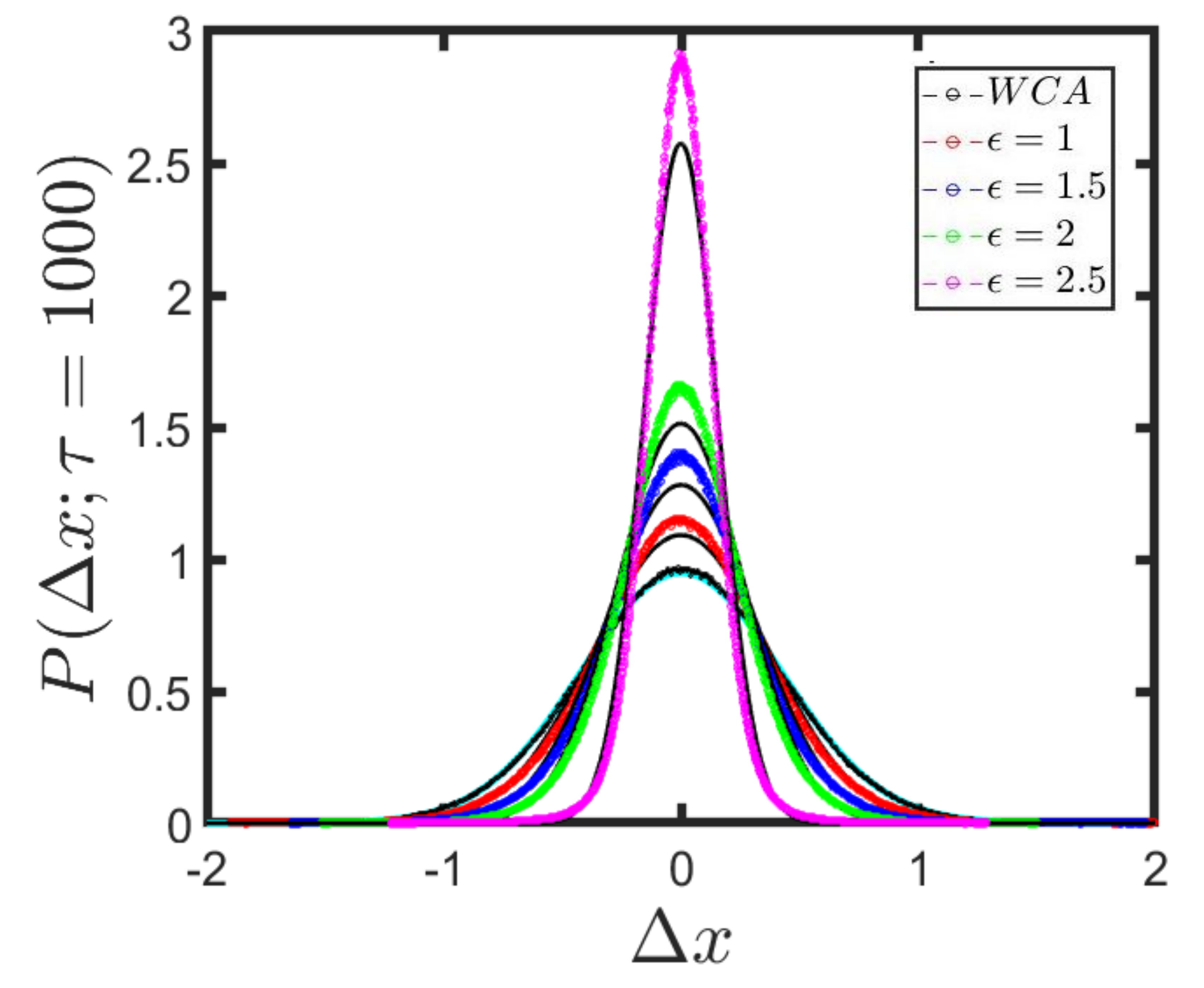} & \includegraphics[width=0.48\linewidth]{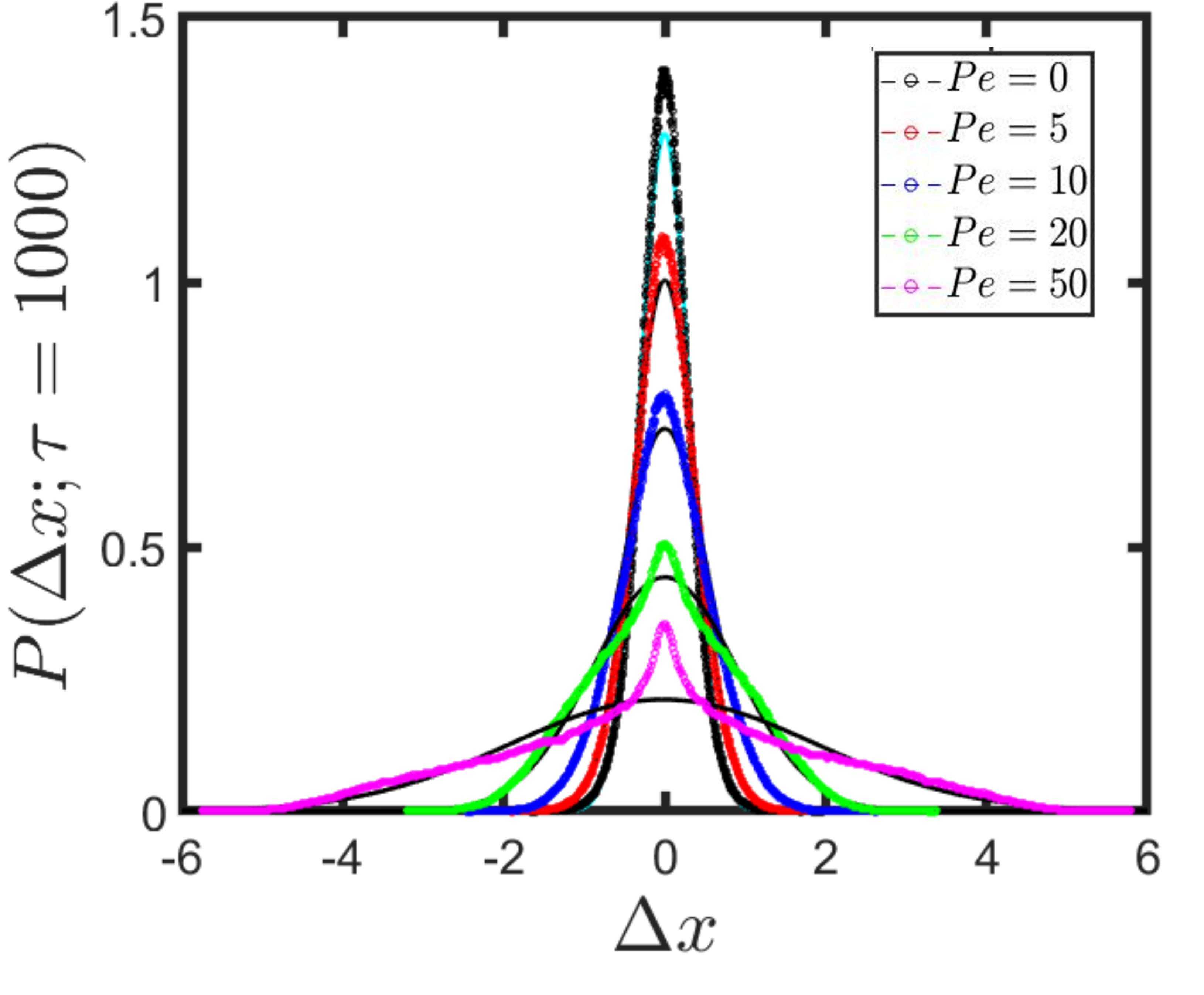} \\
 (a) & (b)
\end{tabular}
 \caption{$\text{P}(\Delta x; \tau)$ of the tracer particle in the polymer grafted cylindrical channel with (a) different $\epsilon$ for $\text{Pe} = 0$ and (b) for different $\text{Pe}$ with $\epsilon = 1.5$. The solid lines (black and cyan) represent the Gaussian fittings.}
 \label{fgr:figure7}
\end{figure*}
%\pagebreak
\clearpage

\noindent {\textbf{Movies}}\\
\noindent The movies illustrate the qualitative difference in the dynamics of the passive and self-propelled tracer particle in the polymer grafted cylindrical channel.

\begin{enumerate}
\item Movie1:
Molecular dynamics simulation of the passive tracer $(\text{Pe} = 0)$ in the polymer grafted cylindrical channel with attractive interaction strength $\epsilon = 1.5$. The tracer particle gets trapped inside the kinks in the local configuration of the grafted polymers, and then as time progresses, the polymers change their configuration, and the tracer escapes (top view).
\item Movie2:
Side view of Movie1. Here we can see that the passive tracer is strongly interacting with the grafted polymers and prefer to stay in the grafted polymeric region. %Molecular dynamics simulation of the passive tracer $(\text{Pe} = 0)$ in the polymer grafted cylindrical channel having attractive interaction strength $\epsilon = 1.5$. The tracer particle gets trapped inside the kinks in the local configuration of the grafted polymers, and then as time progresses, the polymers change their configuration, and the tracer escapes (side view).
\item Movie3:
Molecular dynamics simulation of the self-propelled tracer $(\text{Pe} = 20)$ in the polymer grafted cylindrical channel with attractive interaction strength $\epsilon = 1.5$. The self-propulsion force helps the tracer to escape from the local trap formed by grafted polymers, and the self-propelled tracer undergoes more random paths and explores every part of the channel (top view).
\item Movie4:
Side view of Movie3. It is clearly seen that the active tracer has a tendency to move towards the grafted polymeric zone and prefers to stay close to the wall of the cylinder by following random paths. %Molecular dynamics simulation of the self-propelled tracer $(\text{Pe} = 20)$ in the polymer grafted cylindrical channel with attractive interaction strength $\epsilon = 1.5$. The self-propulsion force helps the tracer to escape from the local trap formed by grafted polymers, and the self-propelled tracer undergoes more random paths and explores every part of the channel (side view).
\end{enumerate}

\providecommand*{\mcitethebibliography}{\thebibliography}
\csname @ifundefined\endcsname{endmcitethebibliography}
{\let\endmcitethebibliography\endthebibliography}{}

\end{document}